\documentclass[twocolumn]{aastex701}

\usepackage{soul}
\usepackage{subfig} %
\usepackage{bm}

\usepackage{booktabs}
\usepackage{multirow}

\newcommand{\astrover}[1]{{#1}}

\newcommand{\ie}{\emph{i.e.}}
\newcommand{\eg}{\emph{e.g.}}
\newcommand{\etc}{\emph{etc.}}

\begin{document}

\title{Probabilistic Data-Driven Modeling of Astrophysical Transients: The Neural Process Family %
for Ultrafast and %
Class-Agnostic
Light Curve Reconstruction with \texttt{NightLANP}
}

\author[orcid=0000-0001-5078-5457,gname=Siddharth, sname=Chaini]{Siddharth Chaini} 
\altaffiliation{NASA FINESST Fellow}
\affiliation{Department of Physics and Astronomy, University of Delaware, Newark, DE 19716, USA}
\affiliation{University of Delaware,
Data Science Institute,
Newark, DE 19716, USA}
\email{chaini@udel.edu}

\author[0000-0003-1953-8727]{Federica B. Bianco}
\affiliation{Department of Physics and Astronomy, University of Delaware, Newark, DE 19716, USA}
\affiliation{Joseph R. Biden, Jr. School of Public Policy and Administration, University of Delaware,
DE 19716,  USA}
\affiliation{University of Delaware,
Data Science Institute,
Newark, DE 19716, USA}
\affiliation{Vera C. Rubin Observatory, Tucson, AZ 85719, USA}
\email{fbianco@udel.edu}

\author[0000-0003-2242-0244]{Ashish Mahabal}
\affiliation{Division of Physics, Mathematics, and Astronomy, California Institute of Technology, Pasadena, CA 91125, USA}
\affiliation{Center for Data Driven Discovery, California Institute of Technology, Pasadena, CA 91125, USA}
\email{aam@astro.caltech.edu}

\begin{abstract}

Astrophysical observations taken from Earth are subject to weather, environmental, and scientific constraints that lead to %
sparse, irregular light curves. 
On the eve of the Vera C. Rubin Observatory Legacy Survey of Space and Time, its massive dataset offers unprecedented opportunities for transient science. 
Yet, a key challenge remains its %
cadence, which will be sparse and irregular across six bands, limiting scientific inference. Interpolation helps mitigate this, 
with Gaussian Processes being the standard, but they struggle with cross-band correlations, require \textit{a priori} kernel specification and must be fit individually to each light curve, hence scaling poorly.
Here, we introduce the neural process family for light curve reconstruction, combining the probabilistic framework of Gaussian Processes with the scalability of deep learning.
By meta-learning on a population of diverse simulated transients,
Attentive Neural Processes shift the bulk of the computational cost to training, enabling rapid, amortized inference with a single, class-agnostic model.
Evaluated on realistic Rubin cadences across 15 transient classes, we show that even an unoptimized, out-of-the-box Attentive Neural Process consistently outperforms all benchmarks---a suite of Gaussian Processes and neural networks---on every tested metric, spanning regression quality, astrophysical feature recovery, and probabilistic calibration. 
Our model interpolates all bands simultaneously in microseconds, over four orders of magnitude faster than the next-best neural benchmark and five %
faster than Gaussian Processes, demonstrating the potential of neural processes and meta-learning for the nightly Rubin alert stream.
Attentive Neural Processes avoid the %
overconfidence of standard neural networks and the underconfidence of Gaussian Processes, delivering sharp, well-calibrated uncertainties. 
This work establishes the neural process family as a scalable, probabilistic foundation for real-time transient science in the Rubin era.

\end{abstract}


\section{Introduction and Background} \label{sec:intro}

Astrophysical observations from ground-based observatories are subject to weather, lunation, and scheduling constraints. Meanwhile, photometric surveys commonly collect observations in alternating filters over time to probe an object's spectral energy distribution (SED), which can help reveal its physical properties like temperature or chemical composition. Thus, astrophysical light curves are characteristically multivariate, sparse, and irregular time series with heteroskedastic noise due to correlation with environmental factors and intrinsic properties of the observed phenomena, posing a challenge for statistical methods.

With the Vera C. Rubin Observatory beginning operations, the Legacy Survey of Space and Time \citep[LSST,][]{ivezicLSSTScienceDrivers2019} will soon deliver millions of transient and variable alerts every night. Conducted in six photometric bands ($ugrizy$) spanning wavelengths from the ultraviolet $(320\ \mathrm{nm})$ to the near-infrared $(1{,}050\ \mathrm{nm})$, LSST will provide an unprecedented dataset for time-domain astronomy.

The LSST survey cadence, however, subject as it is to the additional constraints imposed on the survey by non-time-domain science cases (\eg, large sky area coverage, uniform coverage at data releases, specific cadence in special regions such as the ecliptic, \citealt{ivezicLSSTScienceDrivers2019}), will be sparse even compared to most time-domain surveys. For example, PanSTARRS \citep[$3\Pi$ survey,][]{chambers2019panstarrs1surveys} revisits locations every $\sim5$ days in 6 filters; the Zwicky Transient Facility \citep[ZTF,][]{2019PASP..131a8002B} includes observations of extragalactic fields once per night in its Bright Transient Survey in two or three filters \citep{fremling2020zwicky, perley2020zwicky, rehemtulla2024zwicky};  ATLAS \citep{Tonry_2018} observes in two broad filters with a $\sim2$ day cadence; and ASAS-SN \citep{Shappee_2014} in a single band every 2-3 days. 
Meanwhile, the Wide Fast Deep survey (which is the main survey within the LSST umbrella name and will occupy $\sim$85\% of the observing time) will revisit each location twice in 30 minutes every $\sim3$ nights on average. But these visits will be distributed across the six different filters.\footnote{In reality, the cadence is more complex, including rolling in alternate intensity regions and deep drilling filters. For more information, the reader is encouraged to read \citet{PSTN-056}.}%

These observational challenges are compounded by the sheer scale of the incoming LSST dataset, which is expected to catalog tens of billions of objects over its life, and generate around ten million alerts each night \citep{ivezicLSSTScienceDrivers2019}. This data volume presents a significant computational challenge for traditional resource-intensive analysis workflows which cannot easily scale to process millions of multi-band light curves in real time.

To maximize the scientific return on this deluge of data, time-domain astronomy now increasingly relies on data science and machine learning for diverse tasks \citep[\eg,][]{hambleton2022rubinobservatorylssttransients, collaborationOpportunitiesAIML2026a}; from classification \citep[\eg,][]{sanchez-saezAlertClassificationALeRCE2021, vanroestelZTFSourceClassification2021, qu_photometric_2022, chainiLightCurveClassification2024, sotoSuperphotRealtimeFitting2024, shahORACLERealtimeHierarchical2025} and anomaly detection \citep[\eg,][]{malanchevAnomalyDetectionZwicky2021a, villarDeeplearningApproachLive2021, volnovaExploringUniverseSNAD2024, chainiSearchUnknownUnknowns2025} to parameter estimation \citep[\eg,][]{villarAmortizedBayesianInference2022, karchevSIDErealSupernovaIa2024, simonginiCorecollapseSupernovaParameter2025, brownRapidRobustSimulationbased2026}, population studies \citep[\eg][]{biProbingDiversityType2024, boeskySPLASHRapidHostbased2026}, and systematic searches \citep[\eg,][]{deZwickyTransientFacility2020, toshikageSystematicSearchRapid2024, bommireddyBrokerintegratedAlgorithmElectromagnetic2026}. 

The sparse and irregularly sampled nature of the multi-band LSST light curves limits the performance of models on downstream tasks, including and perhaps especially machine learning approaches, and makes them poorly suited to be coerced into tensor form for input to most neural network architectures. Methods which rely on feature extraction, like classification and regression trees \citep[\eg, \texttt{XGBoost}][]{chen2016xgboost} use features which are, more often than not, extracted independently from each filter \citep[\eg,][\etc]{jainagaAlercebrokerLc_classifierRelease2021, sanchez-saezAlertClassificationALeRCE2021, malanchevLightcurveLightCurve2021, chainiLightCurveClassification2024}. As a result, feature extraction is slow and computationally intensive; the unequal coverage across filters means not all features are equally reliable; and in case of extreme sparsity or absence of these filters, the features cannot be computed, leading to sparse feature catalogs.%

Deep learning has become a popular modern approach, where the allure lies in the model learning feature representations from the light curves automatically during training \citep[\eg,][\etc]{mahabalDeeplearntClassificationLight2017, muthukrishnaRAPIDEarlyClassification2019, mollerSuperNNovaOpensourceFramework2020, shahORACLERealtimeHierarchical2025}. While most traditional deep learning architectures require regular tensors in input, %
attention-based transformers with positional embedding and padding have seen recent success in astrophysics \citep[\eg,][\etc]{donoso-olivaASTROMERTransformerbasedEmbedding2023, cabrera-vivesATATAstronomicalTransformer2024, donoso-olivaAstromer22025} in dealing with irregular tensors.

Recently, machine learning literature has seen a surge in general-purpose Time-Series Foundation Models \citep[TSFMs, \eg,][]{dasDecoderonlyFoundationModel2023, garzaTimeGPT12023, wooUnifiedTrainingUniversal2024, ansariChronosLearningLanguage2024}. These models have 
have shown strong zero-shot performance in fields such as finance and weather, but remain difficult to adapt to the extreme irregularity and sparsity of astronomical data \citep{liStarEmbedBenchmarkingTime2026}. Foundation models specific to time-domain astrophysics models include light-curve-based \citep[\texttt{Astromer2},][]{donoso-olivaAstromer22025} and multi-modal \citep[\texttt{AstroM}$^3$, \texttt{Maven}, \texttt{FALCO};][]{rizhkoAstroM$^3$SelfsupervisedMultimodal2024a, zhangMavenMultimodalFoundation2024, zuoFALCOFoundationModel2025}.

But, of course, these deep learning approaches bind the user to rely on deep neural networks, which, while increasingly explainable, still make interpretation more difficult by limiting the ability to directly `inspect' the light curve, a capability that remains valuable for guiding scientific inference. Another key limitation of most deep learning approaches is the lack of reliable uncertainty estimates and, even when present, these uncertainties are known to suffer from miscalibration \citep{guo2017calibration}. For these reasons, the practice of interpolating, or better, `reconstructing' light curves remains widely used as a preprocessing step 
before direct analysis 
and before applying the above-mentioned machine learning and deep learning methods. 

Gaussian Processes have become the de facto standard for interpolation and reconstruction in time-domain astronomy (TDA), especially after the LSST-focused Kaggle PLAsTiCC challenge \citep{hlozekResultsPhotometricLSST2023}, where the winning solution \citep{boone_avocado_2019} used GPs with a Mat\'ern 3/2 kernel followed by a Light Gradient-Boosting Machine (\texttt{LightGBM}) model for classification. GPs have since become widely used \citep[\eg,][\etc]{villarDeeplearningApproachLive2021, pengKilonovaLightCurveInterpolation2024, qu_photometric_2022, khakpashMultifilterUVInfrared2024, bhardwajPhotometricClassifierTidal2025, magillMALLORNManyArtificial2026, 2026arXiv260403372P}. 
However, GPs suffer from a few drawbacks: they require \textit{a priori} definition of a kernel to model data covariance, and they are computationally expensive since they need to be fit separately to each light curve (see \autoref{subsec:gplim}).%

\begin{figure*}
    \centering
    \includegraphics[width=0.9\linewidth]{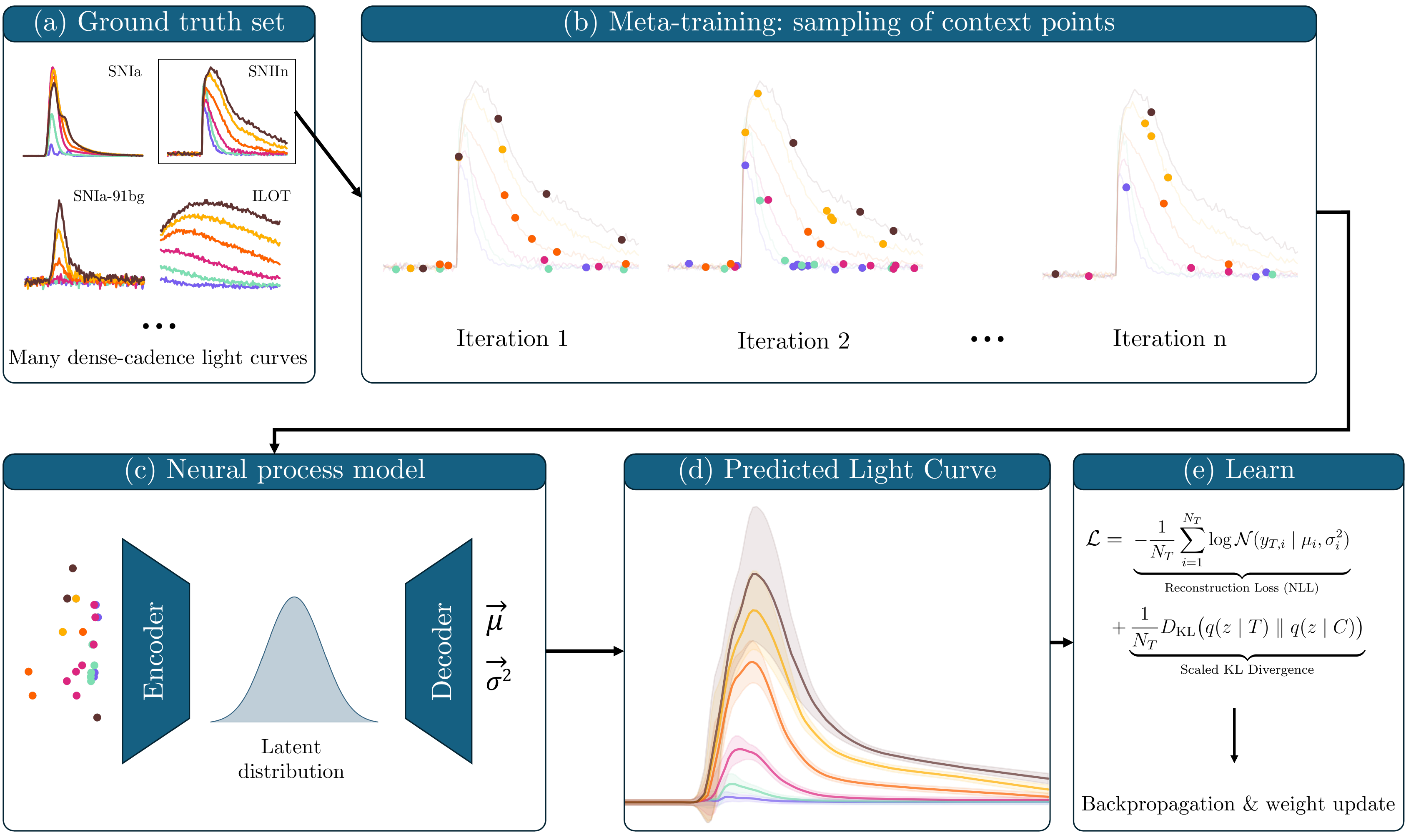}
    \caption{%
    Schematic overview of the meta-learning training procedure for our Attentive Neural Process (ANP) model on light curves. (a) We start with a ground truth set of 15,000 (2- or 10-day cadence) simulated light curves spanning 15 different transient classes (1,000 light curves per class). (b) Meta-training involves choosing a batch of light curves at every iteration and randomly drawing %
    between 12 and 60 points from each light curve, serving as a proxy of the sparse, irregular observations expected from LSST. %
    This forces the model to learn a flexible mapping from any sparse set of observations (context points) to the full light curve (target points), rather than memorizing specific cadence patterns. (c) The neural process model (\autoref{fig:anp-architecture}) consists of an encoder-decoder architecture, and takes only the context points as an input to produce mean and variance predictions, which represents the (d) light curve prediction and the uncertainty, at every point of time. (e) The model is trained to minimize the negative log-likelihood ($\mathcal{L}_\mathrm{NLL}$) loss and KL Divergence via backpropagation, jointly optimizing for both reconstruction accuracy and uncertainty calibration.}
    \label{fig:meta-learning-light-curves}
\end{figure*}

Other approaches %
fit parametric models to recover physical quantities like temperature or bolometric flux \citep{guillochonMOSFiTModularOpen2018, russeil_rainbow_2024}. While physics-driven, these methods impose class-specific physical priors (\eg, a supernova like profile), which limits their flexibility with diverse or unknown transient classes. Additionally, they rely on iterative numerical fitting which may be computationally expensive, limiting their use at an automated survey-scale level.

To model class-agnostic light curves, emergent deep learning methods such as Bayesian neural nets and normalizing flows \citep{demianenko_understanding_2023} have been proposed, alongside %
newer flexible generative models: diffusion-based interpolators that iteratively denoise a latent representation \citep{sasseville_probabilistic_2026, shen_diffusion_2025}, neural-ODE-based methods that parameterize the flux evolution as a differential equation \citep{wuSELDONSupernovaExplosions2026}, 
and transformer-based interpolators that propagate uncertainties through attention layers \citep{yalavarthiTripletformerProbabilisticInterpolation2022, sasseville_probabilistic_2026} -- yet the latter are shown to produce miscalibrated, overconfident uncertainties \citep{sasseville_probabilistic_2026}.
Generally, the latency due to successive forward passes with these methods is also problematic when the goal is to apply them to large datasets in real time, such as the LSST alert stream.

We propose the neural process family \citep{garnelo_conditional_2018, garnelo_neural_2018, kim_attentive_2019} for multiwavelength, sparse and irregular light curve reconstruction. Neural Processes (NPs) are a class of meta-learning neural network models: the same neural network is meta-trained on different subsets of the data. NPs are non-parametric, and rapid at inference. Applications of NPs have emerged, for example, in climate science \citep{scholzSim2RealEnvironmentalNeural2023} but they have seen only limited use in astronomy at the time of writing -- predominantly in AGN modeling \citep{parkInferringBlackHole2021, cvorovic-hajdinjakConditionalNeuralProcess2022, jankovPhotoreverberationMappingQuasars2022, kovacevic_deep_2023, cvorovic-hajdinjakModelingQuasarVariability2024} and isolated applications in exoplanet and gravitational wave studies  \citep{gebhardParameterizingPressureTemperature2024, schuetzRESOLVERareEvent2025}. In these applications, NPs are trained on light curves from a single object class (\ie, that share the same physical processes), and often {used as an offline preprocessing step to simply extract parameters related to that class.

Here, we fundamentally shift this paradigm and specifically propose the use of ANPs (described in \autoref{sec:methods}) as fast, performant and  \emph{universal} transient multi-band light curve interpolators. Unlike previous approaches that assume a predefined, class-specific physical model, our method is entirely morphologically agnostic and meta-learns a shared prior across a diverse population of explosive transients. We show how a single ANP model can perform probabilistic light curve interpolation and extrapolation for different classes of explosive transients (different supernova classes and related phenomena; described in \autoref{sec:data}) without pre-specifying the class during inference. We demonstrate the ANP's superior performance by evaluating its reconstruction under a variety of statistical and physically motivated metrics (described in \autoref{sec:results}).

While we restrict this work to explosive transients, this data-driven framework is naturally extensible to broader classes of transient and variable astronomical sources. We note that \texttt{NightLANP} is not presented here as a production-ready foundation model. Rather, it is a proof-of-concept designed to demonstrate the power of the meta-learning paradigm in solving the bottleneck of astronomical light curve interpolation. By presenting an out-of-the-box architecture that already outperforms established standards, we demonstrate the use of the Neural Process family as a potential foundation for machine learning with light curves. Successful reconstruction methods are expected to enable downstream applications including physical parameter estimation, classification, and anomaly detection, and may be incorporated into other architectures like foundation models; but we leave these for future work and instead highlight here the significant upgrades neural process models can offer.

\section{Methods} \label{sec:methods}

\subsection{Gaussian Processes and their limitations}\label{subsec:gplim}

Traditional supervised machine learning models map a fixed-size input vector to an output vector. For time-domain astronomy, however, we need models that can map inputs of variable lengths (sparse multi-band light curves) to a continuous function, allowing us to query the flux and its uncertainty at any arbitrary time step. 

Gaussian Processes \citep[GPs; \eg,][]{rasmussenGaussianProcessesMachine2008} have seen significant proliferation in astronomy (see \citealt{aigrain2023gaussian}). GPs %
model this continuous function natively by defining a prior over functions with the help of a kernel to model the covariances - for light curves and spectral energy distributions (SEDs), it is this kernel choice that dictates the relationships between flux, time, and wavelength.  Ideally, knowledge of the underlying physical processes would dictate the choice of kernel \citep[\eg,][]{foreman-mackeyFastScalableGaussian2017}. However, in most cases, the kernel is poorly known, or not at all, and the parameters are fit to individual objects. These parameters could be optimized based on a collection of objects of the same class \citep{khakpashMultifilterUVInfrared2024}, yet this requires a robust classification to begin with, while most commonly in applications to large astronomical surveys, a collection of light curves generated by different physical processes is to be modeled (\eg, for classification), such that a physically motivated kernel is not only unknown, but different for different objects. So, in most applications of GPs to survey science, each new light curve is treated individually, and the GP must be fit separately to each, making their use computationally prohibitive. Now, if a new data point is observed for a particular object, the entire GP requires refitting to the complete light curve, where the fitting involves matrix inversion and scales poorly as $\mathcal{O}(N^3)$ with the number of data points $N$ \citep{rasmussenGaussianProcessesMachine2008}.

While faster optimizations exist that leverage tricks like quasi-separable matrices or state space models to speed up computation on GPUs \citep{foreman-mackeyFastScalableGaussian2017, gardnerGPyTorchBlackboxMatrixMatrix2021, rubenzahlScalableGaussianProcesses2026}, they still do not resolve the need for individual fitting for each light curve. This is the fundamental bottleneck that makes GPs impossible to scale for real-time processing, as the LSST alert stream will produce on the order of millions of alerts per night.
This also means that GPs do not carry any `learnt' information across different light curves. The LSST cadence is such that most light curves will have sparse observations compared to the evolutionary time scales of the underlying physics, limiting the information that can be learnt from a single light curve. But what LSST lacks in cadence per object, it will make up for in number; observing many millions of objects. This makes desirable a method that can learn and carry information across different light curves.

\subsection{The Neural Process Family: Attentive Neural Processes}

The above limitations of GPs have motivated the use of neural networks to parameterize a distribution over functions, called the Neural Process \citep[NP; \eg][]{garnelo_conditional_2018, garnelo_neural_2018, kim_attentive_2019, duboisNeuralProcessFamily2020, jhaNeuralProcessFamily2023} family of models, which aim at combining the strengths of GPs and neural networks. 

NPs achieve this through \emph{meta-learning}, where the goal is `learning to learn' \citep{duboisNeuralProcessFamily2020}. Instead of learning from a single dataset (\ie, a single light curve), NPs are trained on a meta-dataset (\ie, a collection of many light curves)\footnote{A meta-dataset refers to a dataset of datasets. If we consider each individual light curve as a dataset containing many data points ${ \{x_i, y_i\}}$, the collection of many light curves is thus a meta-dataset ${ \{x_{i}, y_i\}_j}$ with $i$ denoting the point, and $j$ denoting the dataset.}. At every iteration of the learning cycle (known as an episode, equivalent to a mini-batch in traditional machine learning), a random batch of light curves is drawn from the meta-dataset of light curves (ground truth). The complete light curve forms what is known as a target set, while a small subset of points are sampled randomly from the light curve to form the context set. The goal for an NP model is to use the context set as input, and predict the target set (along with an uncertainty).\footnote{Because of this, the context and target set can be seen as analogous to training and testing sets in traditional deep learning terminology; however note that in our case, a subset is sampled randomly at every episode, and this subset will be different for every episode.} A reconstruction loss is calculated by comparing the model prediction to the target points, and the loss is minimized through backpropagation over many iterations via gradient descent, as in traditional deep learning. Through this procedure, the same NP model learns over a collection of different datasets. A schematic overview of this procedure is shown in \autoref{fig:meta-learning-light-curves}.

Being generative models, NPs comprise an encoder and a decoder. The first neural process model to be developed was the Conditional Neural Process \citep[CNP; ][]{garnelo_conditional_2018}, whose name refers to the fact that it predicts an output \emph{conditioned} on the context set. A CNP is the simplest form of a neural process, and consists of an encoder neural net which maps the vector of context points $(\vec{\bm{x}}_c, \vec{\bm{y}}_c)_i$ to an associated representation vector $\vec{\bm{r}}_i$, which is then averaged into a single vector $\vec{\bm{r}}_*$ representing the complete context. A separate neural net acts as the decoder, and takes as input the context representation $\vec{\bm{r}}_*$, along with the query vector $\vec{\bm{x}}_*$ -- the points where we want to make predictions -- to generate a predicted mean and uncertainty $(\vec{\bm{y}}_{\mu},\ \vec{\bm{y}}_{\sigma^2})$. The Neural Process\footnote{Also referred to as the Latent Neural Process (LNP) in some literature.} model \citep[NP;][]{garnelo_neural_2018} built on the CNP, introducing a parallel latent path to capture global uncertainty via variational inference. This allows the NP to generate multiple coherent realizations of the data, but it still struggled with underfitting.

\astrover{
An extension to the NP is the Attentive Neural Process \citep[ANP;][]{kim_attentive_2019}, which replaces the mean-aggregation with the attention mechanism \citep{vaswaniAttentionAllYou2017}, and is shown in \autoref{fig:anp-architecture}. Like the NP, the ANP consists of two parallel pathways. (1) In the deterministic path, the context points are encoded using self-attention, allowing the model to understand the relationships between the observed data points. Then, a cross-attention mechanism allows the target points to selectively query the most relevant context point depending on the time and band. This means that for a light curve, a missing $r$-band point at peak brightness can strongly attend to an observed $g$-band point from near the peak, while ignoring a $r$-band point from farther before. %
In more detail, labeling the time of observation as $t_i$ and the corresponding observing band $b_i$, in the deterministic path  context pairs $([t_i, b_i], y_i)$ are encoded via self-attention into representations $r_i$. These representations are aggregated via cross-attention using the target locations $( \vec{\bm{t}}_*, \vec{\bm{b}}_*)$ as the query, producing a target-specific representation $\vec{\bm{r}}_*$.
(2) In the latent path self-attention produces representations $s_i$. The context points are encoded into a global summary vector $(\vec{\bm{\mu}}_s, \vec{\bm{\sigma}}_s)$, from which a latent variable $\vec{\bm{z}}$ is sampled. This latent variable $\vec{\bm{z}}$ captures the epistemic uncertainty -- uncertainty related to the model (and the lack of data).   
}

The decoder neural net concatenates the target time/band numbers, the deterministic representation, and the latent sample representation $\vec{\bm{z}}$ to produce a predictive distribution for the flux, characterized by a prediction (mean) $\vec{\bm{y}}_{\mu}$ and an aleatoric uncertainty (variance) $\vec{\bm{y}}_{\sigma^2}$.

For this work, we implemented and tested the use of all three original neural process family flavors: the CNP, the NP, and the ANP. We found that the ANP provides a significant improvement over the CNP and NP, and for the rest of this work, we will only consider ANPs. While the field of neural processes is still actively expanding \citep[\eg,][\etc]{gordonConvolutionalConditionalNeural2020, bruinsmaGaussianNeuralProcess2021, bruinsmaAutoregressiveConditionalNeural2023, ashmanTranslationEquivariantTransformer2024, wangRenyiNeuralProcesses2025, hamadFlowMatchingNeural2025, mortimerIncrementalTransformerNeural2026}, we leave other implementations for future work and use the ANP as a first proof of concept for demonstrating the use of meta-learning via the neural process family for sparse multi-band light curve reconstruction, leading to very rapid inference speeds due to amortized learning. We give more details of our architecture in \autoref{subsec:implementation}.

\begin{figure}
    \centering
    \includegraphics[width=\linewidth]{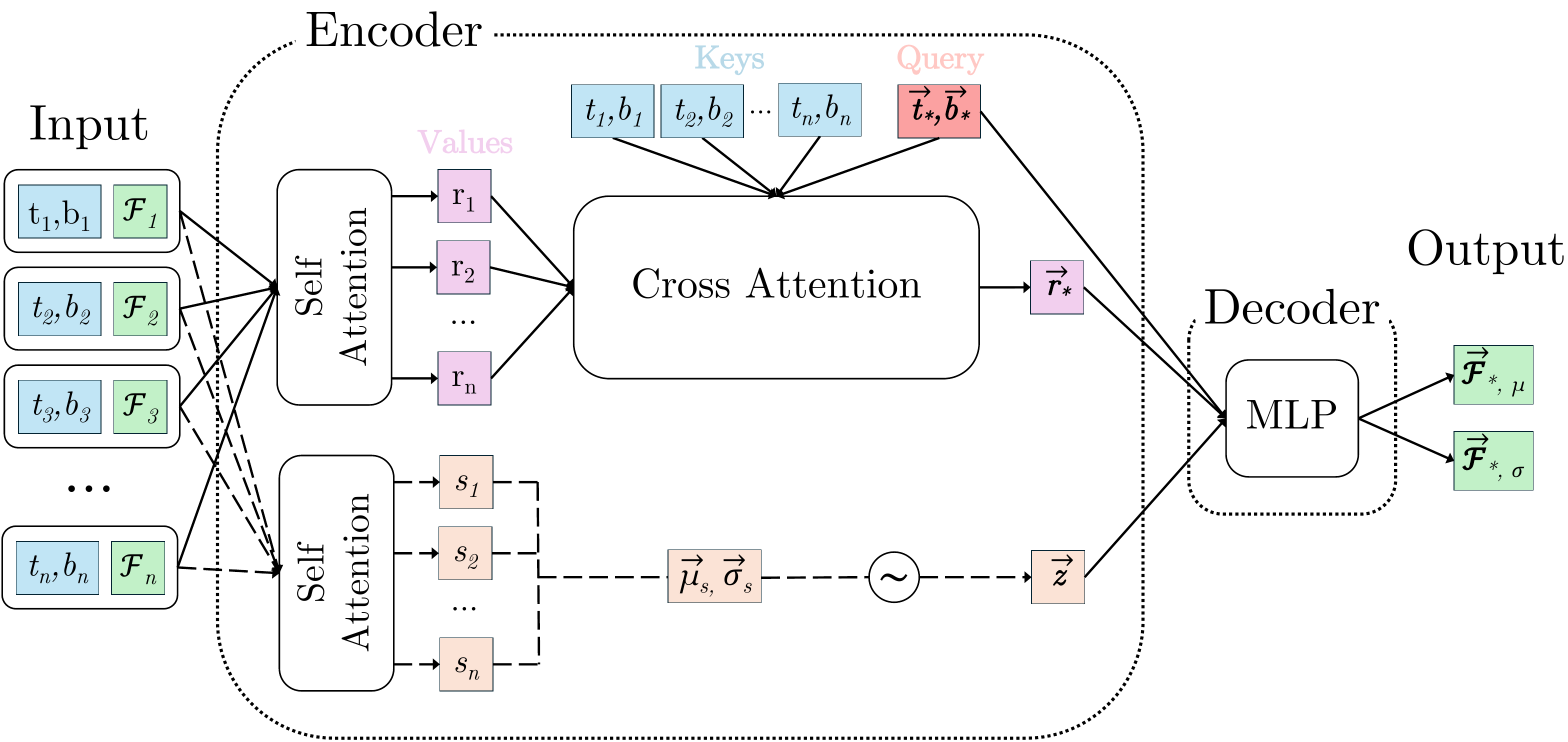}
    \caption{Model architecture of the Attentive Neural Process, adapted from \cite{kim_attentive_2019}. The encoder consists of two parallel pathways, a deterministic and a latent path, described in detail in \autoref{sec:methods}. The decoder concatenates the target location $(\vec{\bm{t}}_*, \vec{\bm{b}}_*)$, the deterministic representation $\vec{\bm{r}}_*$, and the latent sample $\vec{\bm{z}}$, and passes them through an MLP to produce the predictive distribution for $\vec{\bm{\mathcal{F}}}_*$. The cross-attention mechanism allows the model to attend to the most relevant context points for each target. 
    }
    \label{fig:anp-architecture}
\end{figure}

\subsection{Data}\label{sec:data}

Our training data is a dataset of high-cadence simulated transient light curves  \citep{ishidaPERFECTPLASTICCSIM2025}. Most of these light curves %
have observations for all 6 LSST filters over a 2 day cadence (20\% of light curves have a 10-day cadence), and are derived from the PLAsTiCC simulation models \citep{plasticcmodelersLibrariesAmpRecommended2022} for the RESSPECT project \citep{kennamerActiveLearningRESSPECT2020}. Unlike the original PLAsTiCC and other publicly available LSST simulated datasets \citep[\eg,][]{narayanExtendedLSSTAstronomical2023, magillMALLORNManyArtificial2026}, which are all realizations of light curves sampled on a specific implementation of the LSST cadence, this dataset comprises of dense and regular sampling  %
across all six LSST optical filters ($ugrizy$; 80\% of light curves sampled every 2 days, rest are sampled every 10 days), without any observational gaps or weather interruptions. An example of a dense-cadence %
light curve is shown by the crosses in \autoref{fig:sim-to-lsst}.

The full dataset %
consists of nine different models including 14 types of astrophysical transient and variable phenomena.
The models include: 
\begin{enumerate}
\item{three thermonuclear supernova types: SN Ia, Ia91bg, Iax. See \citealt{ruiter2019type} for a review of thermonuclear supernovae including SN Ia and subclasses;}
\item{four core collapse supernova types: SN II, IIP, IIL, IIn. A description of the phenomenology and underlying physics of core collapse supernovae may be found in \citealt{Hamuy:2003xv};}
\item{five stripped envelope supernova subtypes: SN IIb, Ib, Ic, Ibc, and Ic-BL. See, \eg, \citealt{clocchiatti1997sn};}
\item{and two other transients: Calcium-Rich Transients \citep[CART;][]{kasliwal2012calcium} and Intermediate-Luminosity Transients \citep[ILOTs;][]{kashi2017type}}
\end{enumerate}

The dataset is highly imbalanced, with over 300,000 light curves for Type II-Plateau supernovae (SN IIPs), for example, but only 1,589 intermediate luminosity transients (ILOTs). While our pipeline allows us to handle imbalanced datasets (via stratified sampling, as described in \autoref{subsec:adapt-np-astro-training}), we experimented with using hundreds of thousands of light curves and found that convergence took orders of magnitude longer without any drastic improvement on the model. So, we instead work with a class-balanced dataset. In an effort to maximize the information contained in this subset, we selected the longest 1,250 light curves per class, which were then randomly split into 1,000 per class (15,000 total) for training and 250 per class (3,750 total) for testing. We note that the SN IIn type is represented by two distinct PLAsTiCC models (\texttt{NON1ASED.V19\_CC+HostXT} and \texttt{SIMSED.SNIIn-MOSFIT}), which we treat as two separate classes. Our 14 astrophysical phenomena thus correspond to 15 modeling classes, giving 15,000 light curves for training and 3,750 for testing in total.
Full details of our final dataset are given in \autoref{tab:transient_models}. %

\begin{table*}[htpb]
    \centering
        \begin{tabular}{lllccc}
    \toprule
    \toprule
    \textbf{Transient Physics}            & \textbf{Type} & \textbf{PLAsTiCC Model Name}     & \textbf{Full Dataset Size} & \textbf{Train Size} & \textbf{Test Size} \\ 
    \midrule
    
    \multirow{3}{*}{Thermonuclear}     & SN Ia                    & \texttt{SALT2.WFIRST-H17}        & 109,324             & 1,000                & 250                \\
                                       & SN Ia-91bg               & \texttt{SIMSED.SNIa-91bg}        & 11,979              & 1,000                & 250                \\
                                       & SN Iax                   & \texttt{SIMSED.SNIax}            & 31,620              & 1,000                & 250                \\
    \addlinespace
    \midrule[0.1pt]
    
    \multirow{5}{*}{Stripped-Envelope} & SN Ib                    & \texttt{NON1ASED.V19\_CC+HostXT} & 222,578             & 1,000                & 250                \\
                                       & SN Ibc                   & \texttt{SIMSED.SNIbc-MOSFIT}     & 48,508              & 1,000                & 250                \\
                                       & SN Ic                    & \texttt{NON1ASED.V19\_CC+HostXT} & 183,501             & 1,000                & 250                \\
                                       & SN Ic-BL                 & \texttt{NON1ASED.V19\_CC+HostXT} & 36,088              & 1,000                & 250                \\
                                       & SN IIb                   & \texttt{NON1ASED.V19\_CC+HostXT} & 80,902              & 1,000                & 250                \\
    \addlinespace
    \midrule[0.1pt]

    \multirow{5}{*}{Core-Collapse}     & SN II                    & \texttt{SIMSED.SNII-NMF}         & 189,800             & 1,000                & 250                \\
                                       & SN IIL                   & \texttt{NON1ASED.V19\_CC+HostXT} & 184,072             & 1,000                & 250                \\
                                       & SN IIP                   & \texttt{NON1ASED.V19\_CC+HostXT} & 302,829             & 1,000                & 250                \\ 
                                       \cmidrule[0.1pt](l){2-6} 
                                       & \multirow{2}{*}{SN IIn}  & \texttt{NON1ASED.V19\_CC+HostXT} & 51,753              & 1,000                & 250                \\
                                       &                          & \texttt{SIMSED.SNIIn-MOSFIT}     & 14,632              & 1,000                & 250                \\
    \addlinespace
    \midrule[0.1pt]
    
    \multirow{2}{*}{Other}             & CART                     & \texttt{SIMSED.CART-MOSFIT}      & 10,777              & 1,000                & 250                \\
                                       & ILOT                     & \texttt{SIMSED.ILOT-MOSFIT}      & 1,589               & 1,000                & 250                \\
    
    \bottomrule
    \end{tabular}
    \caption{The dataset used in this work, which is a subsampled version of the simulated light curves in \citet{ishidaPERFECTPLASTICCSIM2025}, based on the PLAsTiCC models \citep{plasticcmodelersLibrariesAmpRecommended2022}. Most light curves are sampled every 2 days in the six LSST filters (with 20\% of light curves sampled every 10 days). Note that we treat the two different models for the SNIIn type as different classes.}
    \label{tab:transient_models}
\end{table*}

To evaluate our model, we use the above-mentioned unseen test set of light curves %
, but first generate a degraded version of it from the dense cadence test set of ground-truth light curves. This gives us sparse, irregularly sampled light curves that mimic the %
observational cadence of the main LSST survey. For this, we make use of the official LSST Operations Simulator \citep[OpSims,][]{peteryoachimLsstRubin_simV2612026} version 5.1.1, which simulates the currently planned survey strategy of the LSST survey \citep{PSTN-056}.

We place each object from our test set at a random location (RA \& Dec) within the LSST main survey footprint (Wide Fast Deep). While the OpSims tables include a wide range of other features (\eg, observing conditions, camera parameters), we use them only to impose an LSST-like cadence on top of our dense-simulated light curves, and do not consider factors like galactic latitude. The OpSims table gives us the complete lifetime of LSST observations at the given location; we then choose the start time of our transient to be a random value within the survey lifetime, and extract the dates corresponding to the OpSims observations where our simulated transient would be active. We linearly interpolate the dense-cadence light curves in each filter to retrieve the LSST mock observations. %
An example of an LSST-like realization of a light curve %
is given by the circles in \autoref{fig:sim-to-lsst}. We repeat this procedure a total of ten times for every light curve in our test set %
to later take statistical aggregates across realizations. Each realization results in a light curve having between a minimum of 10 and a maximum of 141 points to form the context set. %

\subsection{Adapting Neural Processes for Light Curves} \label{subsec:adapt-np-astro-training}

\begin{figure}
    \centering
    \includegraphics[width=0.95\linewidth]{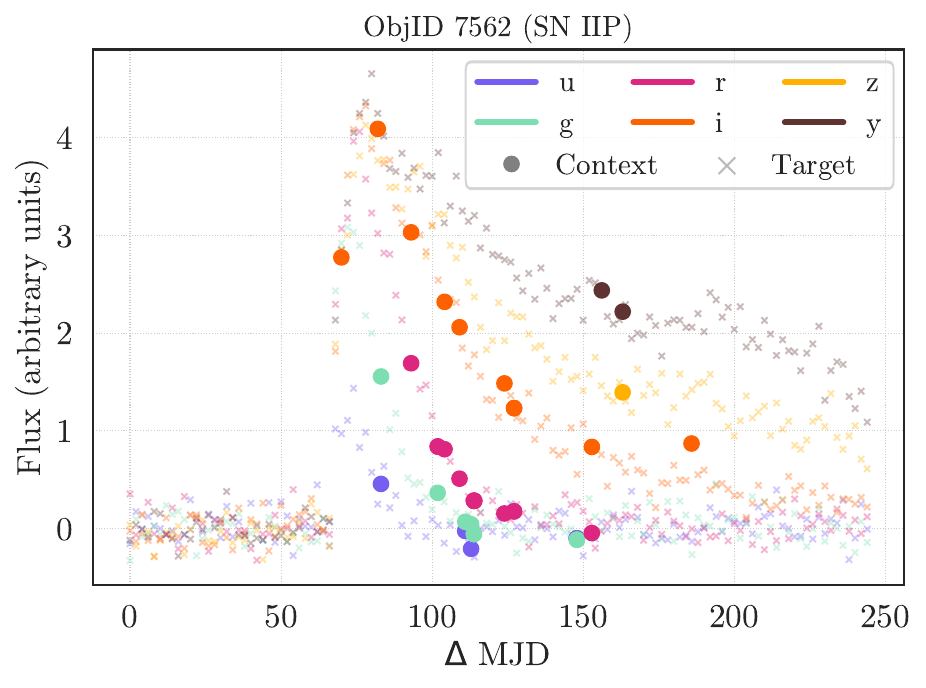}
    \caption{An example of a Type IIP Supernova from our synthetic dataset in the 6 LSST bands ($ugrizy$). The full set of simulated points are denoted by crosses (target points), while a realistic subset that would be observed by LSST is denoted by the circles (context points). Note that this is just a single realization of an LSST cadence based on LSST OpSims v5.1.1. The task of light curve interpolation is to reconstruct the complete multi-band light curve (target points/crosses) from only the sparse observations (context points/circles).}
    \label{fig:sim-to-lsst}
\end{figure}
    
Standard NP implementations in the computer science literature are often highly abstracted to demonstrate GP emulation. To deploy ANPs on real data, we developed a package in Keras 3, \texttt{keras-neural-processes} (Keras 3 \citep{chollet2015keras}, multi-backend across JAX \citep{jax}, PyTorch \citep{pytorch}, and TensorFlow \citep{tensorflow2015-whitepaper}, with ragged-tensor support to handle variable lengths of sparse light curves). \texttt{keras-neural-processes} is general and works with any input of arbitrary dimensions $\vec{\bm{x}}$, to predict any output $\vec{\bm{y}}$ of arbitrary dimensions.

For our astronomical use-case, we have a separate wrapper, \texttt{NightLANP}: Nightly Light curve Analysis with Neural Processes\footnote{\url{https://github.com/sidchaini/NightLANP}}, which utilizes \texttt{keras-neural-processes}\footnote{\url{https://github.com/sidchaini/keras-neural-processes}} for applications to astronomical light curves in particular. 
\texttt{NightLANP} handles all data wrangling via the \texttt{polars} library, and provides an abstraction to move from light curve dataframes to normalized tensors ready to use with neural processes.
\texttt{NightLANP} and \texttt{keras-neural-processes} are intended to be an extensible framework to experiment with different NP architectures, loss functions, and data augmentations. We implement the following four engineering adaptations for use with light curves:

\paragraph{Band-as-coordinate}
Our input is a 2D vector $\vec{\bm{x}} = [t_i, b_i]$, consisting of the Modified Julian date (MJD; $t$) and the categorical band number $b$ for each LSST band $\{u:0, g:1, r:2, i:3, z:4, y:5\}$, while the target is the flux $\vec{\bm{y}} =  [\mathcal{F}_i]$. We pass the band number as an $\vec{\bm{x}}$ coordinate alongside time, rather than have each band as a separate $\vec{\bm{y}}$ channel. Although this choice comes with a computational redundancy when dealing with dense cadence light curves\footnote{For \eg, if a particular date has an observation in each of the six bands, the date needs to be repeated for each band.}, it naturally and elegantly enables us to deal with irregularly sampled light curves, passing a different pair of $[t,b]$ inputs for each different combination of date-band. This is also more data-efficient when dealing with light curves where some filters do not have any observations, making it suitable for LSST-like sparse light curves. While one could explicitly separate the wavelength- from the time-dimension and facilitate the understanding of covariance along the wavelength axis by, \eg, passing band numbers through an embedding layer, for this proof of concept, the ANP automatically learns over the course of its training the cross-band correlations between bands (\eg, how a rise in the $g$-band influences the $r$-band).

\paragraph{Dynamic slicing}
At every episode, once we have sampled a batch of light curves, we randomly crop each full-length light curve to a continuous temporal slice of fixed length (678 points; corresponding to roughly a 57 day window). {We chose this length based on our dataset and architectural choice -- long enough to capture a significant evolution of our transients, but short enough to cover most light curves and run comfortably on our setup (two T4 GPUs). In the few instances when the full light curve has fewer than 678 points, we first sample all available points, and then randomly resample from the slice until we reach 678 points. This cropped light curve slice forms the set of target points, and the context points (to be predicted in this episode) are then randomly sampled from this cropped light curve, drawing from a uniform distribution (see \autoref{fig:sim-to-lsst}). %

The dynamic slicing procedure serves three purposes simultaneously. First, training on a full light curve is memory and computationally prohibitive (especially for our long dense-cadence light curves). Second, it samples from the full irregular light curve to produce target tensors having the same shapes, enabling efficient GPU parallelization. Third, it prevents the model from developing a temporal bias; if we were to crop the light curves to a prescribed region (\eg, centered strictly around peak brightness), the model would learn and assume a phase-centering, making it unsuitable to use on `live' transients for which we may not yet know the peak.

\begin{figure*}
    \centering
    \includegraphics[width=0.95\linewidth]{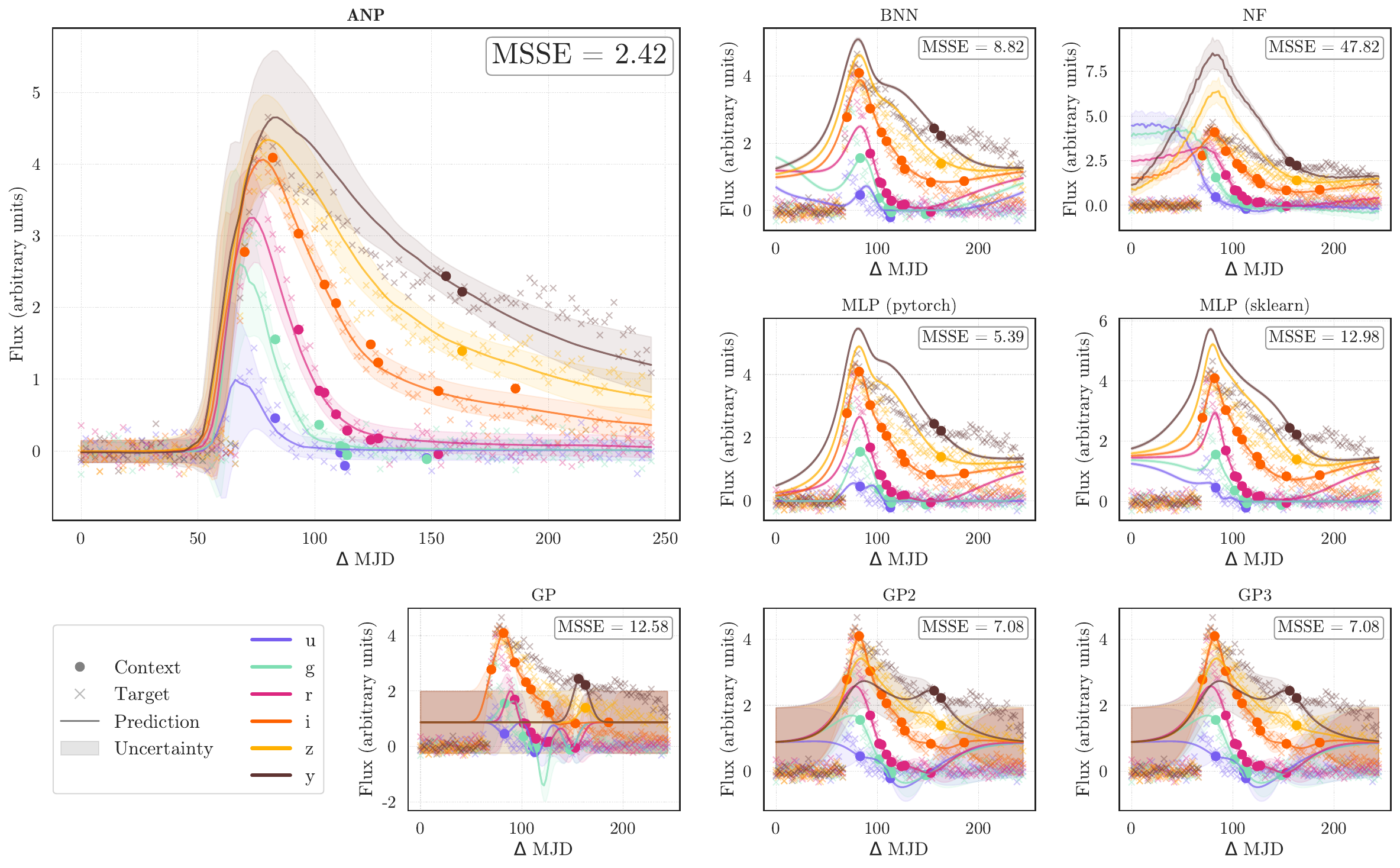}
    \caption{Model predictions for the example Type IIP supernova in \autoref{fig:sim-to-lsst} (ObjID 7562) by our ANP and all seven comparison models considered in this work. The Mean Squared Scaled Error (MSSE; lower is better) of each prediction is indicated for each. The predictions (means) are shown as solid lines, and $\pm$1$\sigma$ uncertainty bands (shaded regions) are shown (when available) for each of the six LSST bands $ugrizy$, overlaid on the context points (circles) and target points (crosses). Our ANP result is shown in the top-left panel (enlarged). The remaining panels show the same light curve as interpolated by the seven benchmark models (implemented from \citealt{demianenko_understanding_2023}, see \autoref{sec:results}): Bayesian Neural Network (BNN), Normalizing Flow (NF), Multilayer Perceptron (MLP) implemented in \texttt{pytorch} and \texttt{sklearn}, and a Gaussian Process with three different variants of kernels (GP: \texttt{C(1.0) * RBF([1.0, 1.0]) + WhiteKernel()}, GP2: \texttt{C(1.0)*RBF([1.0, 1.0]) + Matern() + WhiteKernel()}, GP3: \texttt{C(1.0)*Matern() * RBF([1, 1]) + Matern() + WhiteKernel()}). 
    The ANP outperforms all the models with the lowest MSSE (2.42), roughly capturing the rise and decay. The GP variants (GP2: 7.08, GP3: 7.08) provide the next-best reconstructions but produce sharper features and are unable to generate accurate predictions in bands with fewer points ($u$ and $y$), with correspondingly large uncertainties. %
    All models except ANP rapidly regress to the mean before and after the first context point and, in some cases (\eg, GP), in large gaps between observations. The ANP has learnt the behavior of light curves as a collection during meta-learning (see \autoref{subsec:procedure}), allowing it to accurately predict the trend with high confidence (low uncertainty).
    }
    \label{fig:models-lc-compare}
\end{figure*}

\begin{figure*}
    \centering
    \includegraphics[width=0.95\linewidth]{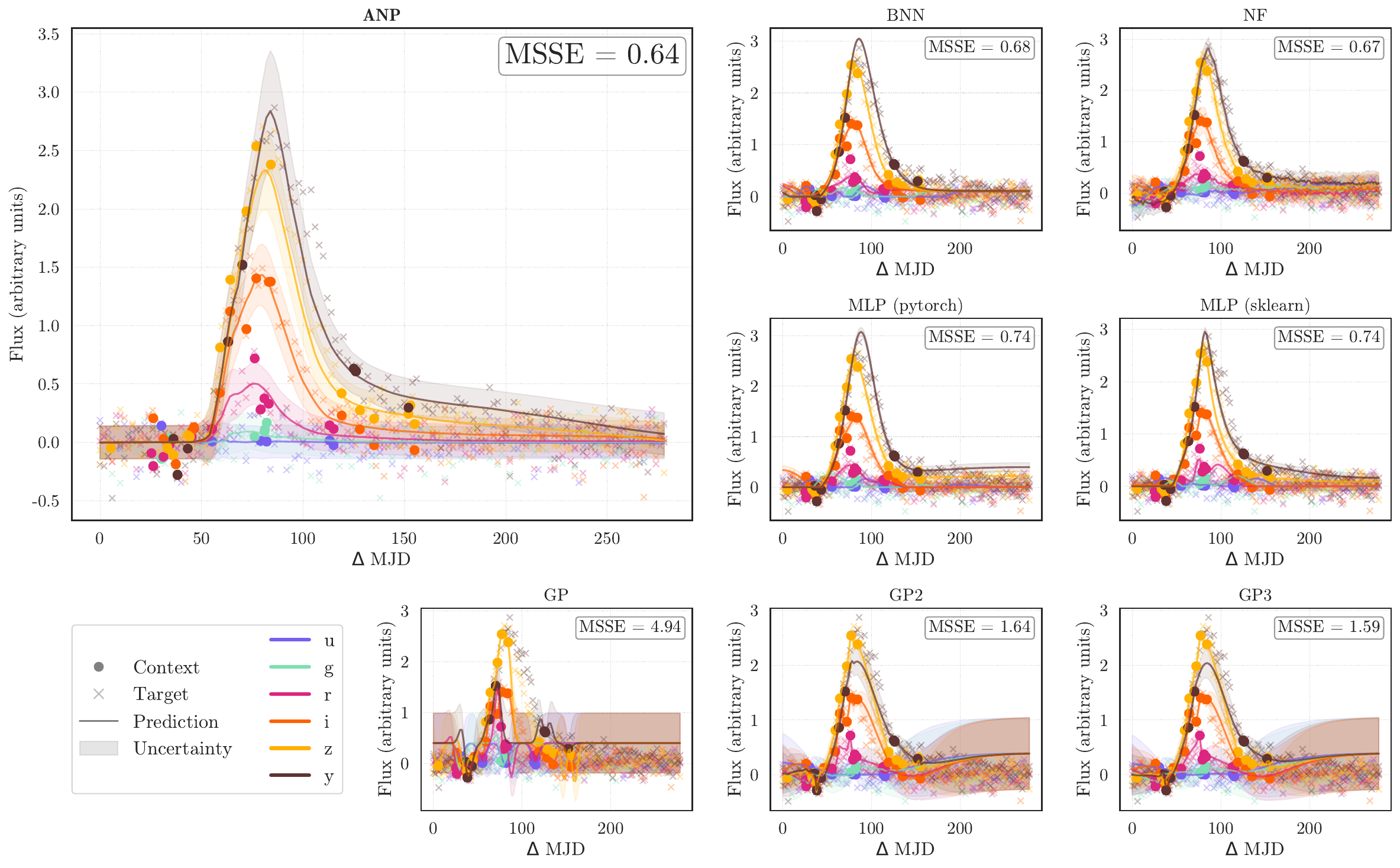}
    \caption{As \autoref{fig:models-lc-compare} for a Type Iax Supernova (ObjID 35623). The predictions from the ANP are the most accurate, but in this case, where more context points are available both before and after peak, other neural network models (BNN, NF, MLPs; see \autoref{sec:results} and \autoref{fig:models-lc-compare}) produce comparable results.
    }
    \label{fig:models-lc-compare2}
\end{figure*}

\paragraph{Temporal re-zeroing}
While ANPs do not enforce translational equivariance implicitly, we implement an on-the-fly augmentation where after sampling and slicing our target light curve, we subtract the start time of the slice, re-zeroing the temporal slice for every batch. This forces the ANP to learn from relative times only, allowing for use on partial (incomplete) light curves without any peak-alignment. We also note that because of this, the same light curve results in different target slices with different starting points over the course of training.

\paragraph{Stratified sampling}
To handle class-imbalanced datasets, we implement stratified sampling. In every training loop, the batch is sampled with roughly equal class representation in each batch. While this may lead to over-representation from the minority classes, it ensures the majority classes do not dominate the loss. Combined with the randomization in dynamic slicing and context selection, the ANP effectively never sees the same temporal configuration twice, making it a suitable data augmentation strategy.

\subsubsection{Observational Uncertainties with Neural Processes}

A note on uncertainties: the ANP results presented in this work do not incorporate observational flux uncertainties from the light curves: each context point is encoded as $(\vec{\bm{x}}, \vec{\bm{y}})_i = ([t_i, b_i], \mathcal{F}_i)$. Our ANP implementation is, however, naturally extensible to allow the use of these uncertainties: $(\vec{\bm{x}}, \vec{\bm{y}})_i = ([t_i, b_i], [\mathcal{F}_i, e_{i}])$, where $e$ is the observational uncertainty in flux for that point. This approach is a smaller departure from the original implementations of ANP and, when we compare our model performance with that of existing models (\autoref{sec:results}), aligns our input with those of the Multilayer Perceptrons (MLP) and Bayesian Neural Network in \citet{demianenko_understanding_2023}. Additionally, $e_i$ can then be incorporated into the predictive likelihood, similar to how a GP treats per-point noise via the kernel. In future work (Chaini et al., in preparation), we will focus on ANP architectures that enable the incorporation of these aleatoric uncertainties in input and in the loss function.

\subsection{Training Procedure} \label{subsec:procedure}
During training, we pass the full dense-cadence training set to the ANP model, which performs `dynamic slicing' internally, sampling a set of context points and target points, where the context points are a small fraction (between 1.8\% and 8.8\%) of the target points.\footnote{In practice, we sample the context-set length from 10 fixed options rather than from a continuous uniform distribution. This allows for efficient GPU speedup via JIT compilation, since each time a previously-unseen length is encountered, a re-compilation is required.} 

While this ANP model outperformed all benchmark models in all metrics we set forth to test, we noted that (like other models) it was struggling at forecasting tasks from rise-only observations, which are critical in astronomy (\eg, for follow-up \cite{hambleton2022rubinobservatorylssttransients}). So, we additionally train a separate forecasting-version of the ANP, where during training, 50\% of the context points in each episode are randomly cropped to contain pre-peak points only. We use this version only for the forecasting task, and report its performance on the forecasting predictions of the peak in \autoref{subsec:results-peak}.

\subsection{ANP Implementation} \label{subsec:implementation}

Our ANP model consists of a deterministic encoder, a latent encoder, and a decoder. The model is implemented in Keras 3 within our package \texttt{NightLANP} and \texttt{keras-neural-processes}. The deterministic encoder processes points through a 4-layer MLP, while the latent encoder and decoder each use a 2-layer MLP, with 128 ReLU-activated hidden units in each layer, except the output layers, which are all unactivated. The deterministic path uses self- and cross-attention with 8 heads, while the latent path uses mean-pooling to parameterize a 128-dimensional global latent distribution.

We trained the model for 700,000 episodes with a batch size of 64 light curves per episode.\footnote{Training was performed till sufficient convergence; it is possible more optimal convergence parameters could be used.}  The training objective is to minimize the negative log-likelihood loss combined with the KL divergence between the variational posterior and prior latent distributions, using the Adam optimizer with a constant learning rate of $10^{-4}$. The full set of hyperparameters and configuration files used for the runs reported here are available in our open-source \texttt{NightLANP} package.

We note that we did not perform an extensive architectural search or ablation studies. Because our goal is to validate the effectiveness of NPs for multiband time-domain astronomy, rather than to engineer a production-ready model, we deliberately report the first configuration that converged stably. As we will demonstrate in detail in \autoref{sec:results}, 
this configuration already systematically outperforms a robust set of existing models---established benchmark models in the astrophysical literature \citep{demianenko_understanding_2023}---which highlights how naturally suited the meta-learning approach is for handling irregular light curves. We expect further tuning would yield additional improvements.

\section{ANP performance} \label{sec:results}
The objective of the model is to reconstruct the complete, multi-band dense light curve --- the \emph{target} set (shown as crosses in \autoref{fig:sim-to-lsst}) relying solely on the sparse context points representing the LSST-like cadence (shown as circles in \autoref{fig:sim-to-lsst}).

We compare our model (ANP) against seven benchmark reconstruction methods, using the implementation in \texttt{fulu} \citep{demianenko_understanding_2023}. These include a Bayesian Neural Network (BNN), Normalizing Flows (NF), two Multilayer Perceptron (MLP) implementations - in PyTorch and scikit-learn, and a Gaussian Process (GP) with three distinct kernel combinations.

All models take the context set as their input, which is an LSST-like realization of a simulated transient light curve (see \autoref{sec:data}), and make a prediction of what the complete light curve should look like -- two example reconstructions comparing ANP with the other models are given in \autoref{fig:models-lc-compare} and \autoref{fig:models-lc-compare2}. %

We use a broad range of metrics to compare prediction performance between the ANP and other benchmark models -- regression quality with and without uncertainty, recovery of astrophysical features near the peak, probabilistic metrics and uncertainty quantification metrics. Each of these metrics is described in detail in \autoref{subsec:results-regression}-\ref{subsec:results-probabilistic}. A summary of the results of the different metrics is provided %
in \autoref{fig:winrate comparison} (see also \autoref{app:summary} \autoref{tab:metrics_summary}) which shows the win rate (inset percentage) for each model: the percentage of light curves for which the specified model is the best performer. 
The ANP is \emph{overwhelmingly} the best model for reconstruction according every tested metric, with an over 80\% win rate on all statistical metrics, and still outperforming the benchmarks in the evaluation of the physical properties of the reconstructed light curves.

\begin{figure*}
    \centering
    \includegraphics[width=\linewidth]{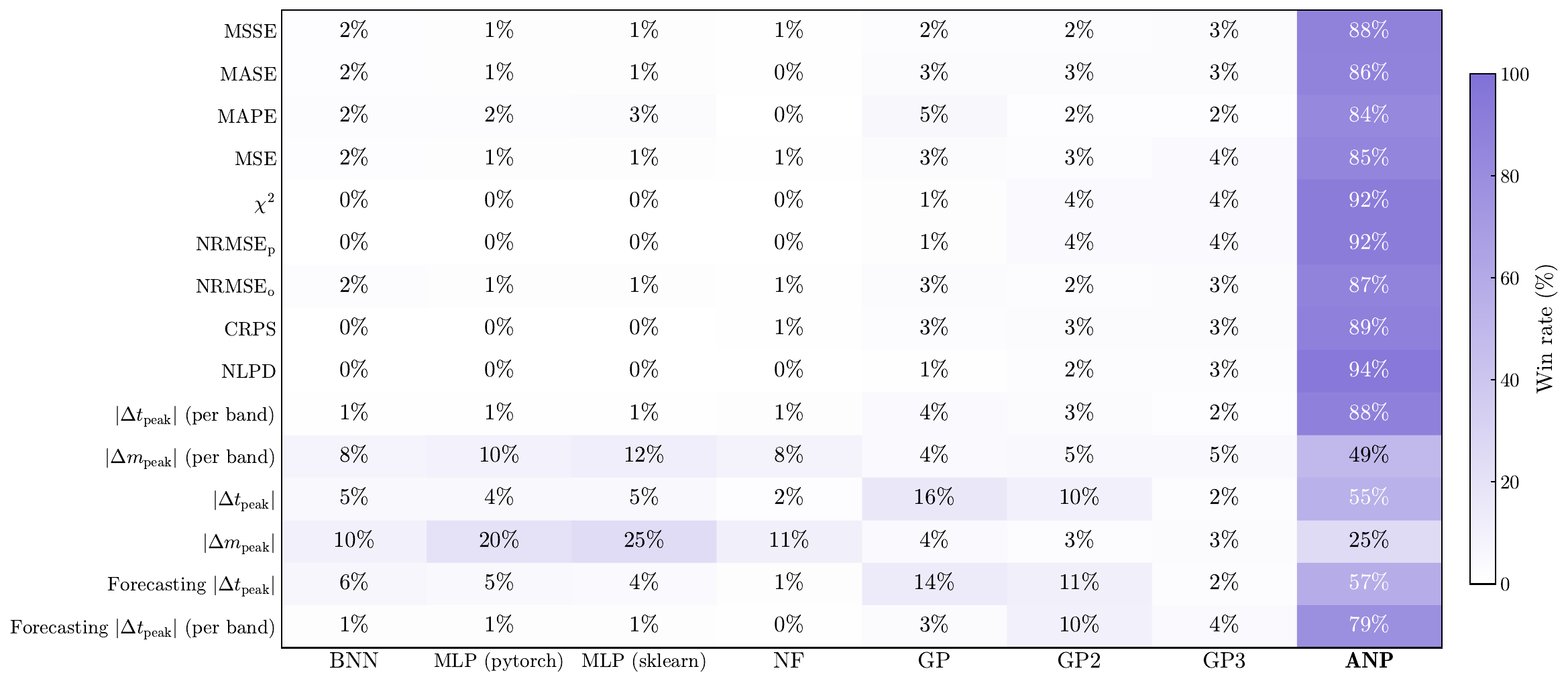}
    \caption{Summary of the performance of the ANP and seven benchmark models for all metrics considered in this work (see \autoref{sec:results}). The inset percentage (win rate) specifies the percentage of light curves for which the specified model is the best performer as per that metric (higher is better). ANP is overwhelmingly the best model for all reconstruction metrics (top 9 rows) and when considering the prediction of time and magnitude of peak, globally or per band (next four rows, see \autoref{subsec:results-peak}). Only in the global magnitude prediction do MLP models compete with ANP, but ANP is still at least as good as the next-best model (see also \autoref{app:summary} \autoref{tab:metrics_summary}).}
    \label{fig:winrate comparison}
\end{figure*}

\begin{figure*}
    \centering
    \includegraphics[width=0.8\linewidth]{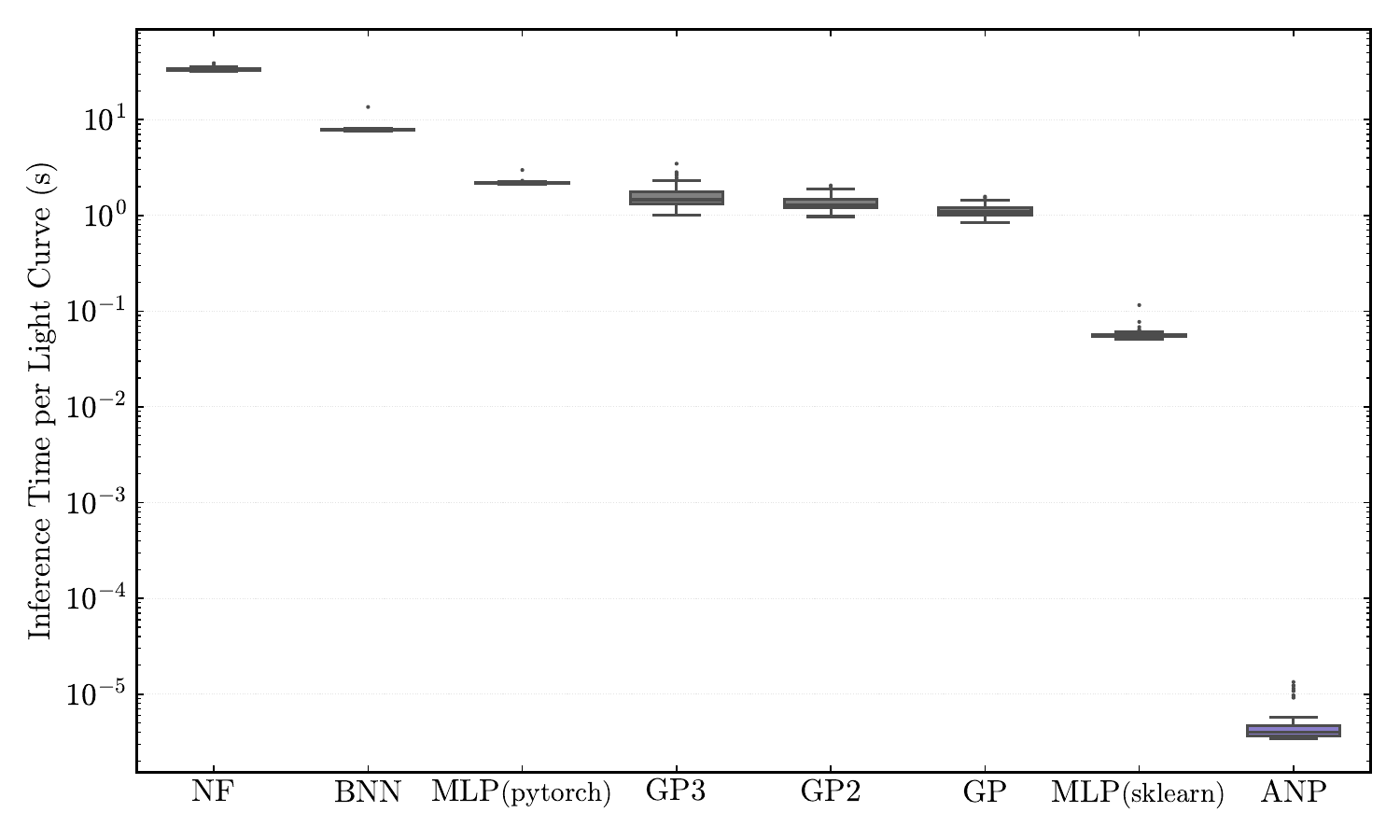}
    \caption{Distribution of per light curve inference times (in seconds, log $y$-axis) for ANP and all seven benchmark models considered in this work (\autoref{sec:results}), excluding training time (lower is better). %
    ANP achieves $\sim$10$^{-6}$~s/light curve -- four orders of magnitude faster than the next-best neural benchmark (MLP) and over five orders of magnitude faster than GPs.}
    \label{fig:timing comparison}
\end{figure*}

Before we dive into each of these, however, we first contrast the computational complexity of the different models; something of paramount consideration when handling the massive LSST dataset.

\subsection{Timing and Computational Complexity} \label{subsec:results-time}

An important distinction between the benchmark reconstruction models and the ANP is inherent to the meta-learning framework: ANPs have a computationally expensive meta-training phase, when they train on the training set described in \autoref{sec:data}. %
Conversely, the other benchmark models %
must be individually initialized and fit from scratch to the context points of every single light curve in the test set. %

In this amortized inference model, no fitting or training is required when ANPs see a new light curve, which makes the model orders of magnitude faster than the other benchmark models, as shown in \autoref{fig:timing comparison}. Evaluated over 80 runs processing parallelized batches of 500 light curves, the ANP achieves an effective inference time of $(4.92 \pm 2.38) \times 10^{-6}$~seconds per light curve on two NVIDIA T4 GPUs.%
\footnote{The T4 is freely available through online services like Google Colab.} This makes inference with the ANP four orders of magnitude faster than the next-best model (MLP; $\sim 5.7 \times 10^{-2}$~seconds), and over five orders of magnitude faster than GPs ($\sim 1.2 $~seconds), the current standard.

\begin{figure*}[!t]
    \centering
    \subfloat[]{
        \label{subfig:}
        \includegraphics[height=0.275\linewidth]{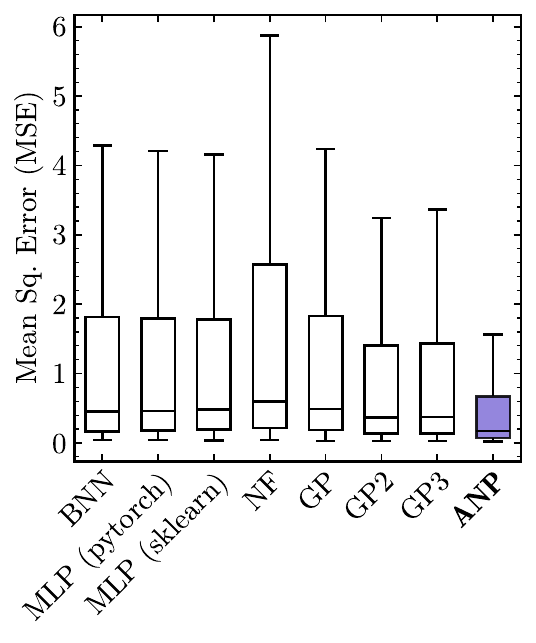}
    }
    \subfloat[]{
        \label{subfig:}
        \includegraphics[height=0.275\linewidth]{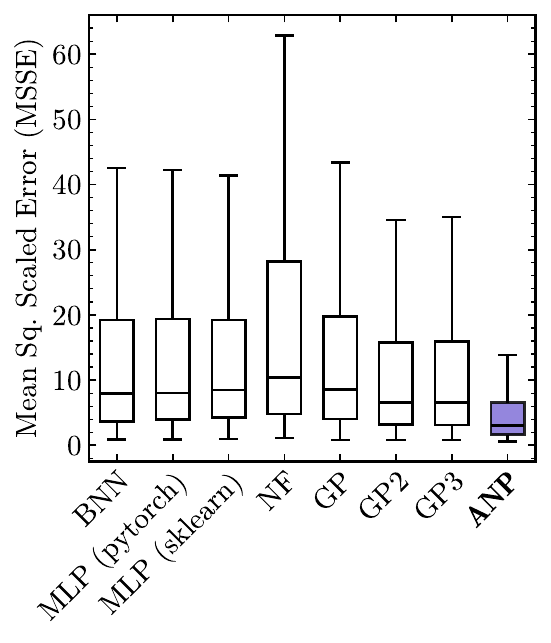}
    }
    \subfloat[]{
        \label{subfig:}
        \includegraphics[height=0.275\linewidth]{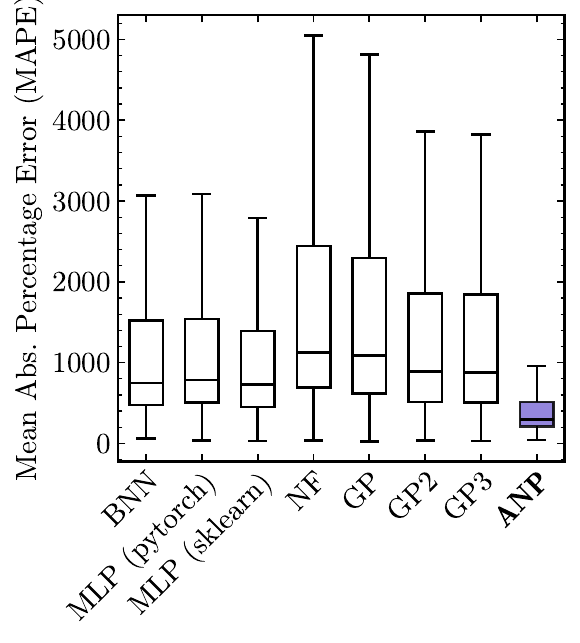}
    }
    \subfloat[]{
        \label{subfig:}
        \includegraphics[height=0.275\linewidth]{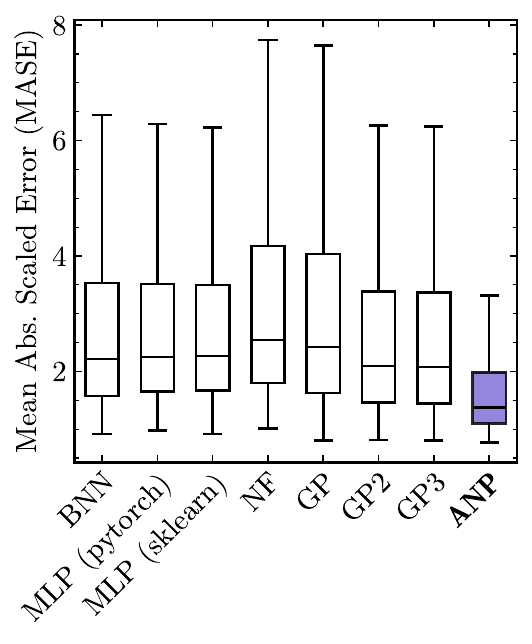}
    }
\caption{Results I: Distribution of four regression quality metrics (\autoref{subsec:results-regression}) for the unseen 3,750 light curves (15 transient classes, 250 objects each) across all eight reconstruction models (lower is better). Each light curve includes context points for 10 realizations based on the expected LSST cadence. (a) Mean Squared Error (MSE) is computed per object across all bands, while (b) Mean Squared Scaled Error (MSSE), (c) Mean Absolute Percentage Error (MAPE), and (d) Mean Absolute Scaled Error (MASE) are computed per object per band and averaged across bands. All metrics are finally median-aggregated over the 10 cadence realizations to mitigate pseudoreplication \citep{hurlbertPseudoreplicationDesignEcological1984} due to the different cadence realizations. Boxes span the interquartile range (IQR); whiskers extend to 1.5$\times$IQR. Outliers outside this range have been omitted for clarity. The ANP (purple, filled) outperforms all models, achieving the lowest median on all four metrics as well as the most compact distribution.}
    \label{fig:result-regression}
\end{figure*}
Training the ANP on the dataset described in \autoref{sec:data} took 331 minutes. Consider the size of the LSST data releases: the first Data Release \citep[DR1,][]{RTN-011} is expected to contain $\sim3\times10^{10}$ light curves \citep{DMTN-135}. Running the ANP as a first-pass interpolation model before any downstream tasks will take $\sim$ 1 hour.

\begin{figure*}
    \centering
    \subfloat[]{
        \label{subfig:}
        \includegraphics[height=0.275\linewidth]{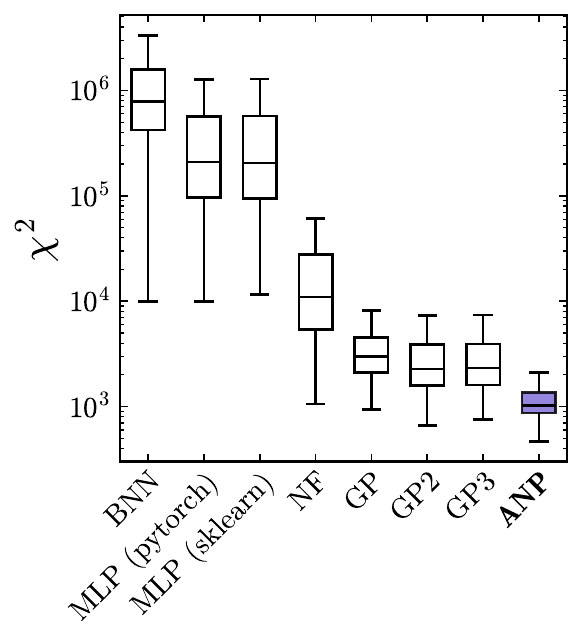}
    }
    \subfloat[]{
        \label{subfig:}
        \includegraphics[height=0.275\linewidth]{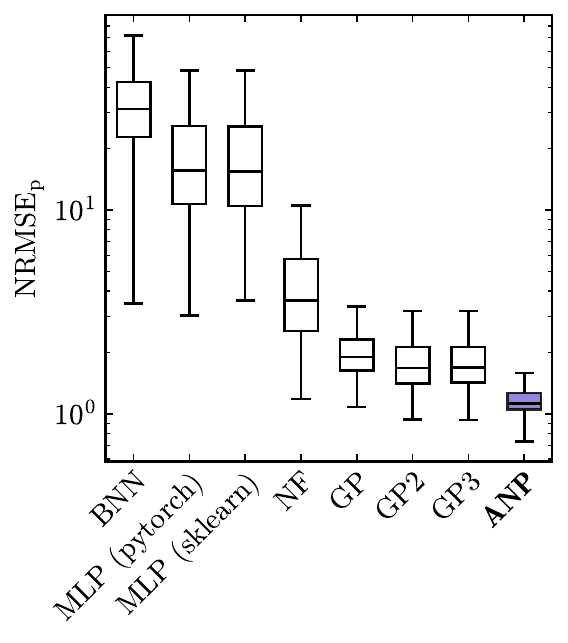}
    }
    \subfloat[]{
        \label{subfig:}
        \includegraphics[height=0.275\linewidth]{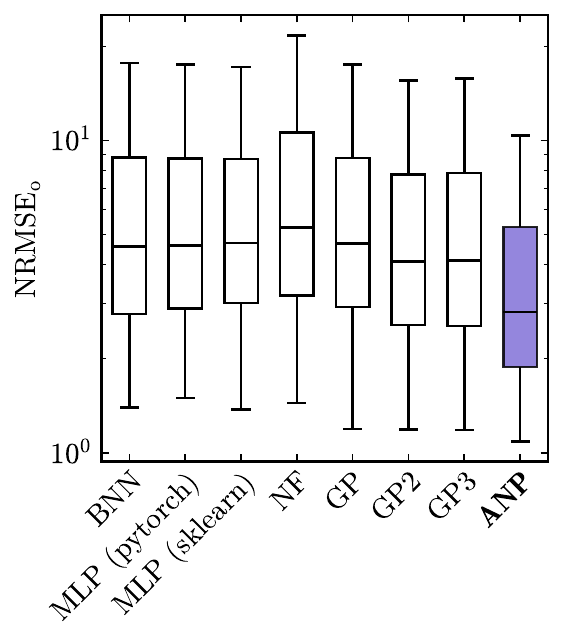}
    }
    \caption{Results II: As \autoref{fig:result-regression} for three uncertainty-aware regression metrics (see \autoref{subsec:results-uncertainty}) that evaluate regression quality while taking the uncertainties into consideration (lower is better). Note that all the metrics ($y$-axis) here have been plotted on a logarithmic scale. The ANP (purple, filled) achieves the lowest median and most compact distribution on all three metrics. In particular, the entire distribution of ANP results outperforms all BNN and MLP results as measured by $\chi^2$ and $\mathrm{NRMSE}_\mathrm{p}$ (see \autoref{subsec:results-uncertainty} for a description of the metrics).}
    \label{fig:result-uncertainty}
\end{figure*}
However, consider the application of the %
ANP to nightly alerts once trained -- a single night is expected to produce $\sim$7 million alerts, most of which will be new points for a light curve that is currently evolving. The ANP can ingest all of these and make a prediction in well under a minute, making it possible to rapidly update each light curve prediction as new alerts arrive in real time, allowing for a nightly evolving set of model predictions. This makes the ANP capable of running `online' on this evolving physical system, and generating uncertainty-aware representations that evolve over time.}

While this speed is a key strength of the ANP, it is not its only strength - we will now show that not only is the ANP exceptionally fast, but also overwhelmingly the best at reconstruction compared to all models we tested.

\subsection{Reconstruction Quality} \label{subsec:reconstruction-performance}

To quantitatively assess the reconstruction quality of the different models, we follow the procedure of \citet{demianenko_understanding_2023} with slight modifications. We evaluate model predictions given the context against the hidden ground-truth dense-cadence test light curves. 
As described in \autoref{subsec:procedure}, for each of the 3,750 test light curves, we generated 10 independent LSST-like realizations from their dense cadence light curves. The goal is to truly compare errors due to the inherent model, and mitigate the extra source of error (the cadence). We note that some realizations, however, result in `unlucky' observational cadences, \eg, where context points are sampled only at the beginning or at the tail end of the light curve. In these cases, models generally perform poorly, causing large reconstruction errors. Producing multiple realizations, we get a fairer assessment of the performance based on the nature of the light curve.
However, these realizations represent different observational subsets of the same underlying physical light curve. Treating them as completely independent data points would lead to incorrect statistical comparisons, due to the pseudoreplication effect \citep{hurlbertPseudoreplicationDesignEcological1984}.
For this reason, we calculate the median metric across the 10 realizations for each of the test light curves before comparing reconstruction metrics.

In the following we measure the performance of ANP against that of the baseline models for a set of traditional and physics-motivated metrics commonly used in astrophysics (\autoref{fig:result-regression}-\ref{fig:forecast-time-peak}), and a set of probabilistic metrics (\autoref{fig:result-picpmpiw}-\ref{fig:result-prob}). We use box and whiskers plots for each metric in \autoref{fig:result-regression}-\ref{fig:result-picpmpiw} and \autoref{fig:result-prob} to show the performance scores across all test light curves for each model. In all these plots, the ANP is highlighted (in purple for the regular interpolation version, green for the forecasting version).

\subsubsection{Standard Regression Metrics} \label{subsec:results-regression}
Standard regression metrics evaluate the accuracy of the models' point predictions $(\vec{\bm{\mathcal{F}}}_{\mu})$ against the ground truth flux values. While all our models predict a mean and an uncertainty, for this class of metrics, only the mean is evaluated against the true value,  ignoring the uncertainty estimate.

We compute the (a) Mean Squared Error (MSE), (b) Mean Squared Scaled Error (MSSE), (c) Mean Absolute Percentage Error (MAPE) and (d) Mean Absolute Scaled Error (MASE), plotted in \autoref{fig:result-regression}. We compute these per object per band and average across bands - this ensures we treat all bands equally, and consistency across bands is rewarded. 

While MSE provides an unnormalized metric in flux units, it inherently overweights the intrinsically bright transients and the brighter bands. For this reason, we introduced the scaled form of the errors, the MSSE (MASE), which divides the MSE (MAE) by the mean variation of the consecutive truth flux values, making it more suitable for comparing the light curves of objects and bands across different flux ranges. We also include MAPE for completeness, though we note that this metric would be dominated by the errors of near-zero flux values.

The ANP (\autoref{fig:result-regression}, purple, filled) outperforms all benchmark models, achieving the lowest median as well as the most compact distribution on all four metrics.

\subsubsection{Uncertainty-aware regression Metrics} \label{subsec:results-uncertainty}
While standard regression metrics evaluate the accuracy of the point prediction $(\vec{\bm{\mathcal{F}}}_{\mu})$, they do not assess the model's error estimates $(\vec{\bm{\mathcal{F}}}_{\sigma^2})$. In time-domain astronomy, especially in the context of follow-up optimization, `overconfidently wrong' will have adverse effects, causing a waste of limited and costly resources and potential loss of scientific discoveries. So, we also evaluate uncertainty-aware regression metrics that explicitly penalize predictions when the true flux falls far outside the predicted error bounds. 

We compute three such metrics, (a) $\chi^2$ goodness of fit, (b) Normalized Root Mean Squared Error evaluated with respect to the predicted error ($\text{NRMSE}_{\text{p}}$) and (c) Normalized Root Mean Squared Error evaluated with respect to the observed error ($\text{NRMSE}_{\text{o}}$), which have been plotted in \autoref{fig:result-uncertainty}. %
The observed errors here are the errors on the inputs obtained from the SNANA PLAsTiCC models.%

We find that the ANP achieves the lowest median and the most compact distribution across all three metrics. The difference is much more drastic for predicted error (b), compared to observed errors (c). Other neural models (BNN, MLP, NF) are orders of magnitude worse, indicating that they are highly overconfident, a known property of neural networks \citep[\eg,][]{guo2017calibration}. While the GP variants perform better than the standard neural networks, they still exhibit errors roughly twice as large as the ANP. We will return to the discussion on uncertainty and confidence in \autoref{subsec:results-probabilistic}, when we talk about probability theory-driven metrics.

\begin{figure*}
    \centering
    \subfloat[]{
        \label{subfig:}
        \includegraphics[height=0.27\linewidth]{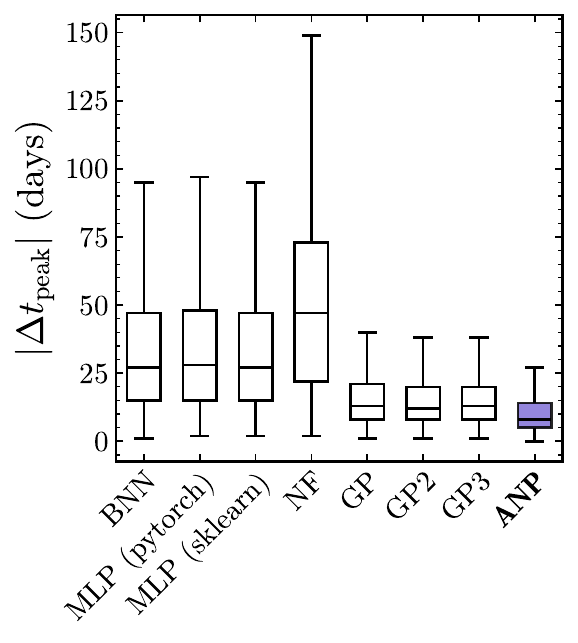}
    }
    \subfloat[]{
        \label{subfig:}
        \includegraphics[height=0.27\linewidth]{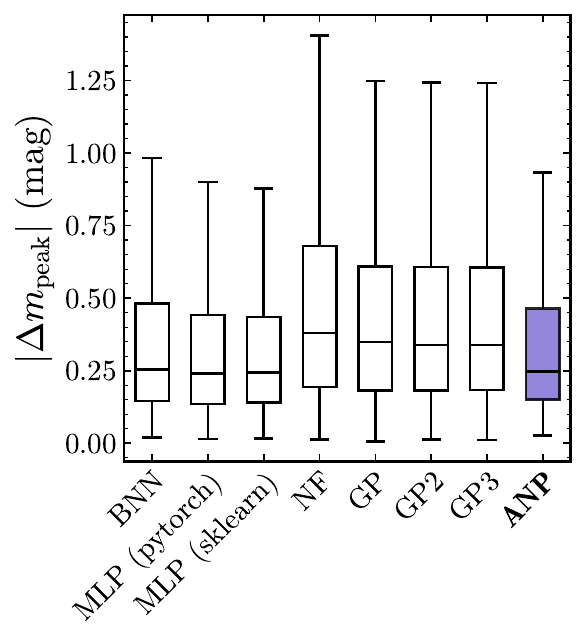}
    }
    \subfloat[]{
        \label{subfig:}
        \includegraphics[height=0.27\linewidth]{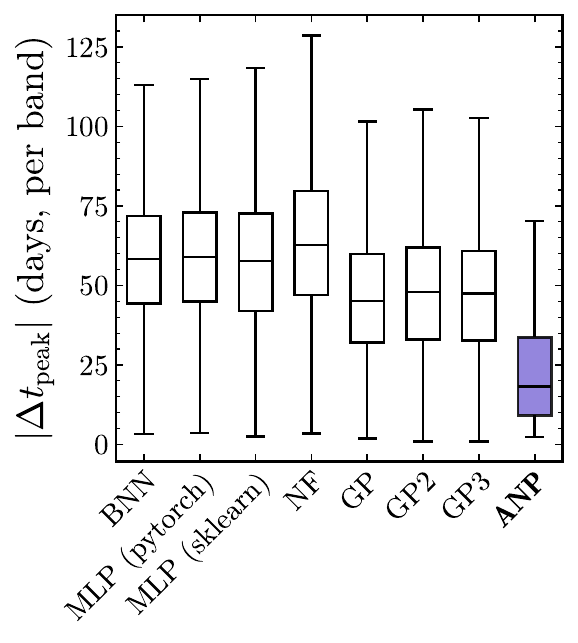}
    }
        \subfloat[]{
        \label{subfig:}
        \includegraphics[height=0.27\linewidth]{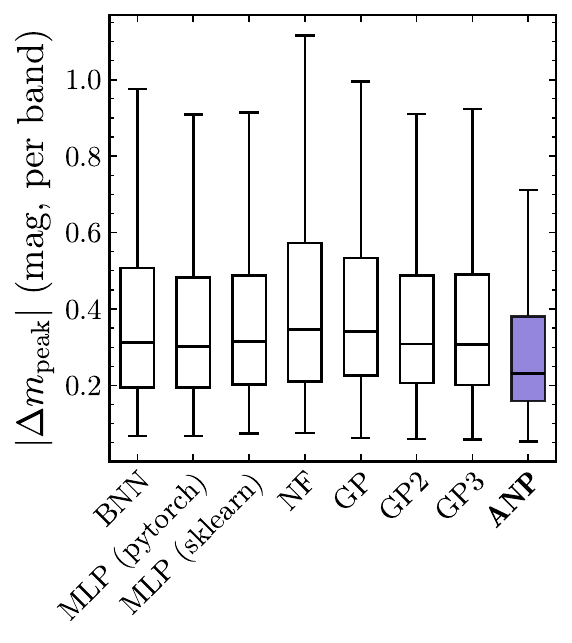}
    }
    \caption{Results III: Distribution of four peak estimation metrics (see \autoref{subsec:results-peak}), evaluating how accurately each model recovers the time and magnitude of peak brightness (lower is better). %
    Like in \autoref{fig:result-regression}, all metrics are median-aggregated over 10 cadence realizations. (a) Absolute peak-time error ($|\Delta t_\mathrm{peak}|$, in days) for the global light curve peak computed across all bands. (b) Absolute peak magnitude error ($|\Delta m_\mathrm{peak}|$, in luptitudes) for the global light curve peak computed across all bands. (c) Absolute per-band peak-time error, averaged across bands. (d) Absolute per-band peak magnitude error. ANP (purple, filled) is consistently the top performer.}
    \label{fig:result-peaks}
\end{figure*}

\begin{figure*}
    \centering
    \subfloat[]{
        \label{subfig:}
        \includegraphics[height=0.27\linewidth]{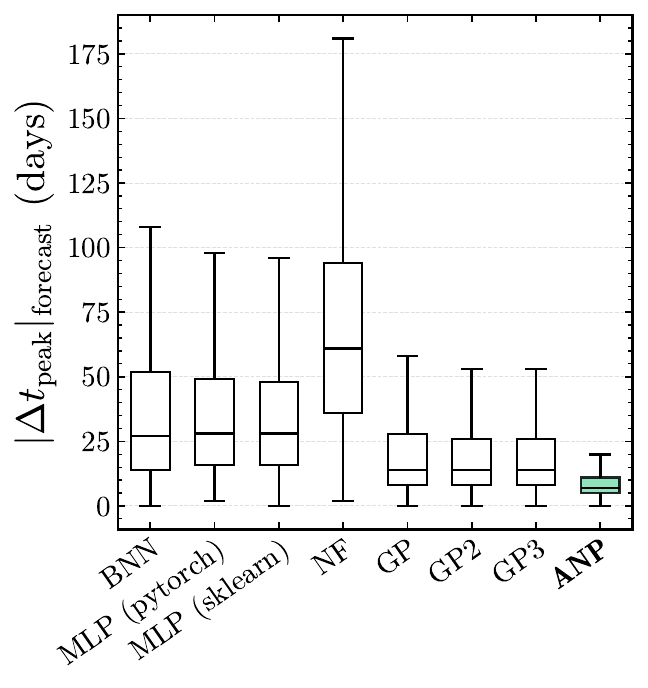}
    }
    \subfloat[]{
        \label{subfig:}
        \includegraphics[height=0.27\linewidth]{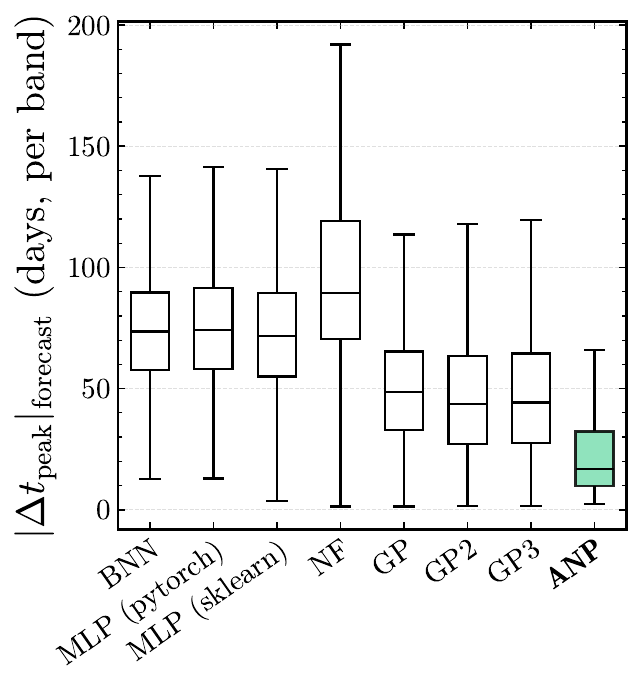}
    }
    \caption{Results IV: Distribution of peak-time errors like \autoref{fig:result-peaks} (a) and (c), but for the task of forecasting: when the context consist of points observed pre-peak from the rising part of the light curve (Forecasting $|\Delta t_\mathrm{peak}|$, in days) (see \autoref{subsec:results-peak}). Here, we use a separate forecasting version of the ANP (green, filled), where during training, 50\% of the context points in each episode were randomly cropped to contain pre-peak points only. Accurate forecasting of peak-time may help inform follow-up recommendations. (a) Absolute peak-time error for the global light curve. (b) Absolute per-band peak-time error, averaged across bands. The forecasting-trained ANP outperforms all other models.}
    \label{fig:forecast-time-peak}
\end{figure*}

\begin{figure}
    \centering
    \includegraphics[height=0.75\linewidth]{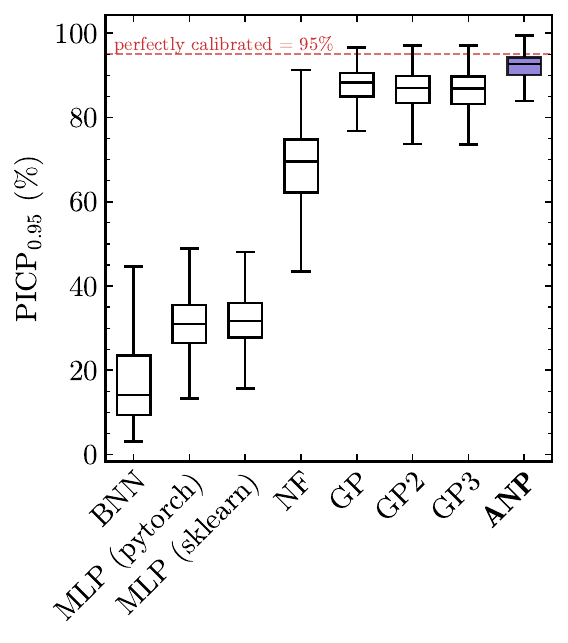}
    
    \includegraphics[height=0.75\linewidth]{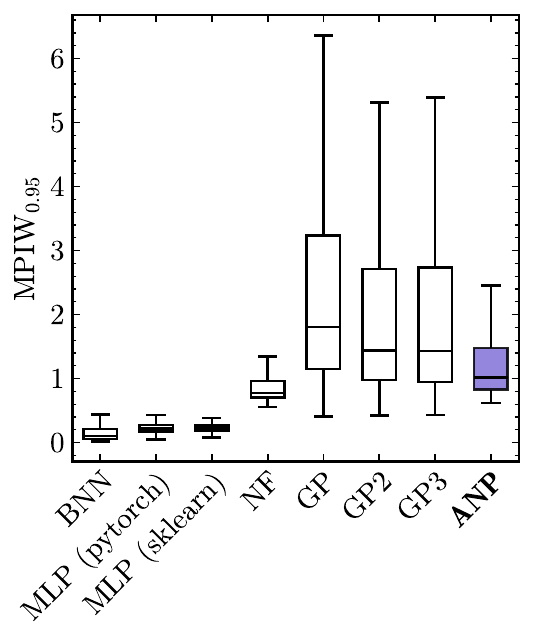}

    \caption{Results V: Distribution of two probabilistic evaluation metrics (see \autoref{subsec:results-probabilistic}). Like in \autoref{fig:result-regression}, all metrics are median-aggregated over 10 cadence realizations. (a) Prediction Interval Coverage Probability (PICP) at the 95\% confidence level -- the fraction of true flux values falling within the 95th-percentile of the predicted values distribution. (b) Mean Prediction Interval Width (MPIW) at the 95\% level, measuring the average width of the confidence interval. Reading these results together, not only are the ANP predictions more accurate, but the other neural models (\eg, BNN, MLPs, and NF) tend to overestimate their confidence relative to ANP (high precision and low accuracy).}
    \label{fig:result-picpmpiw}
\end{figure}

\subsubsection{Peak-estimation Metrics} \label{subsec:results-peak}
In the context of transient science, knowledge of the time and magnitude of peak brightness is directly relevant for planning follow-up spectroscopy, classifying transients, estimating distances for cosmology, correctly assigning spectral phases, \etc{}
For this reason, we also evaluate a physically motivated set of peak-estimation metrics: the absolute error in the predicted time of peak $|\Delta t_{\text{peak}}|$ (in days), and the absolute error in the predicted peak magnitude $|\Delta m_{\text{peak}}|$. Magnitudes are calculated as asinh magnitudes or `luptitudes' \citep{luptonModifiedMagnitudeSystem1999} from the flux, to gracefully handle near-zero and negative flux values. We use a fixed zero-point (27.5) for all filters. Because the flux units are arbitrary, the resulting magnitude values are also arbitrary -- but the \emph{difference} in magnitude values are still consistent across the different models we are comparing.

We calculate $|\Delta t_{\text{peak}}|$ and $|\Delta m_{\text{peak}}|$ from the full predicted light curve with two approaches. First, a global approach: we measure the absolute maximum flux across all six filters and take the absolute difference between its time (magnitude) and the maximum time (magnitude) measured on the ground-truth light curve. Second, a per-band approach: we calculate the maximum flux in each of the six filters separately, measure the absolute difference between its time (magnitude) and the true maximum time (magnitude) for that band, and then averaged across bands. The per-band metric rewards correct color estimates, and thus measures the accuracy of the estimated Spectral Energy Distribution (SED). %

Finally, it is critical to distinguish between two separate astrophysical applications: peak prediction for inference and peak forecasting for follow-up. Let us begin with the former.}

\paragraph{Peak prediction for light curves observed throughout}
\label{subsec:results-peakprediction}

In this scenario, the transients have been observed (sparsely) for their entire duration (\eg, using the light curves from an LSST Data Release catalog). Knowing the time of peak enables, for example, timing spectral time series, aligning models, and aligning observed events in time for statistical comparison. The results from the four metrics described in \autoref{subsec:results-peak} are plotted in \autoref{fig:result-peaks}. The ANP is the best performer for all metrics. For peak-time prediction (panels a and c), the ANP outperforms the benchmarks, providing an accurate reconstruction of the time evolution and the SED; this is especially apparent for the per-band approach (panel c). For peak magnitude prediction (panels b and d), other neural models (BNN and MLPs) perform comparably to ANP, but are less accurate in locating the peak in time.

\paragraph{Forecasting peak-time from rise-only contexts} \label{subsec:results-forecasting}

Now we turn our attention to forecasting peak-time when only observations pre-peak are available, a highly important application of light curve reconstruction methods that enable effective and efficient follow up. 

When we train the ANP providing context points throughout the entire light curve, we unsurprisingly find it to be inaccurate at forecasting when only pre-peak observations are available; we thus implemented a separate training strategy for forecasting (see \autoref{subsec:procedure}) where 50\% of the context points during training are pre-peak. This version of our ANP then outperforms all our benchmark models for both the global and per-band metric (\autoref{fig:forecast-time-peak}). Note that the GP is the next-best for global peak forecasting, but we argue that this is an artifact of the tendency of GPs to regress to the baseline in absence of observations. That is to say: GPs will generally predict a drop in magnitude shortly after the last observed data point, which, by design, is near peak in our simulations. Generally, all models struggle to forecast accurate peak-times, but ANPs trained for forecasting are a promising prospect. We leave more tests of this training schema to future work.

\subsubsection{Probabilistic evaluation metrics} \label{subsec:results-probabilistic}

To comprehensively evaluate the probabilistic nature of our forecasts, we adopt the evaluation paradigm proposed by \cite{gneitingProbabilisticForecastsCalibration2007}: probabilistic forecasts should aim to ``maximize the sharpness of the predictive distributions subject to calibration''.
In this context, ``calibration'' refers to the statistical consistency between the distributional forecasts and the distribution of observations (for example, does the true flux value actually lie within the 95\% confidence interval of the prediction 95\% of the time?). On the other hand, ``sharpness'' refers to the compactness of the predictive distributions (for example, 
how tight are the error bars in the prediction?). We prefer models with smaller error bars (sharper), but only if those error bars contain the true values (well calibrated).

We use the Prediction Interval Coverage Probability (PICP) and the Mean Prediction Interval Width (MPIW) to quantify calibration and sharpness, respectively. If we assume our models' output to be a Gaussian distribution characterized by $(\vec{\mathcal{F}}_{\mu},\vec{\mathcal{F}}_{\sigma^2})$, we can map each confidence percentage to an interval based on the Gaussian. The PICP thus measures the empirical fraction of ground truth flux values that fall within the predicted uncertainty intervals, while the MPIW calculates the average width of these predicted intervals.

For all models, we calculate the PICP and MPIW at the 95\% confidence interval -- $\text{PICP}_{0.95}$ calculates the fraction of true flux values falling within the 95th-percentile of the predicted values distribution, while $\text{MPIW}_{0.95}$ measures the average width of the 95\% confidence interval. From \autoref{fig:result-picpmpiw} (upper), we see that the ANP predictions are the best calibrated, and closest to the perfect calibration value of 95\%. Meanwhile from \autoref{fig:result-picpmpiw} (lower), we see that the other neural models (BNN, MLP, and NF) have the smallest prediction intervals; however, from the PICP we know that these are all instances of the model being confidently wrong, confirming that they overestimate their confidence. Reading these together, we infer that the ANP is the best calibrated model at the 95\% confidence interval.

\begin{figure}
    \centering
    \includegraphics[width=1\linewidth]{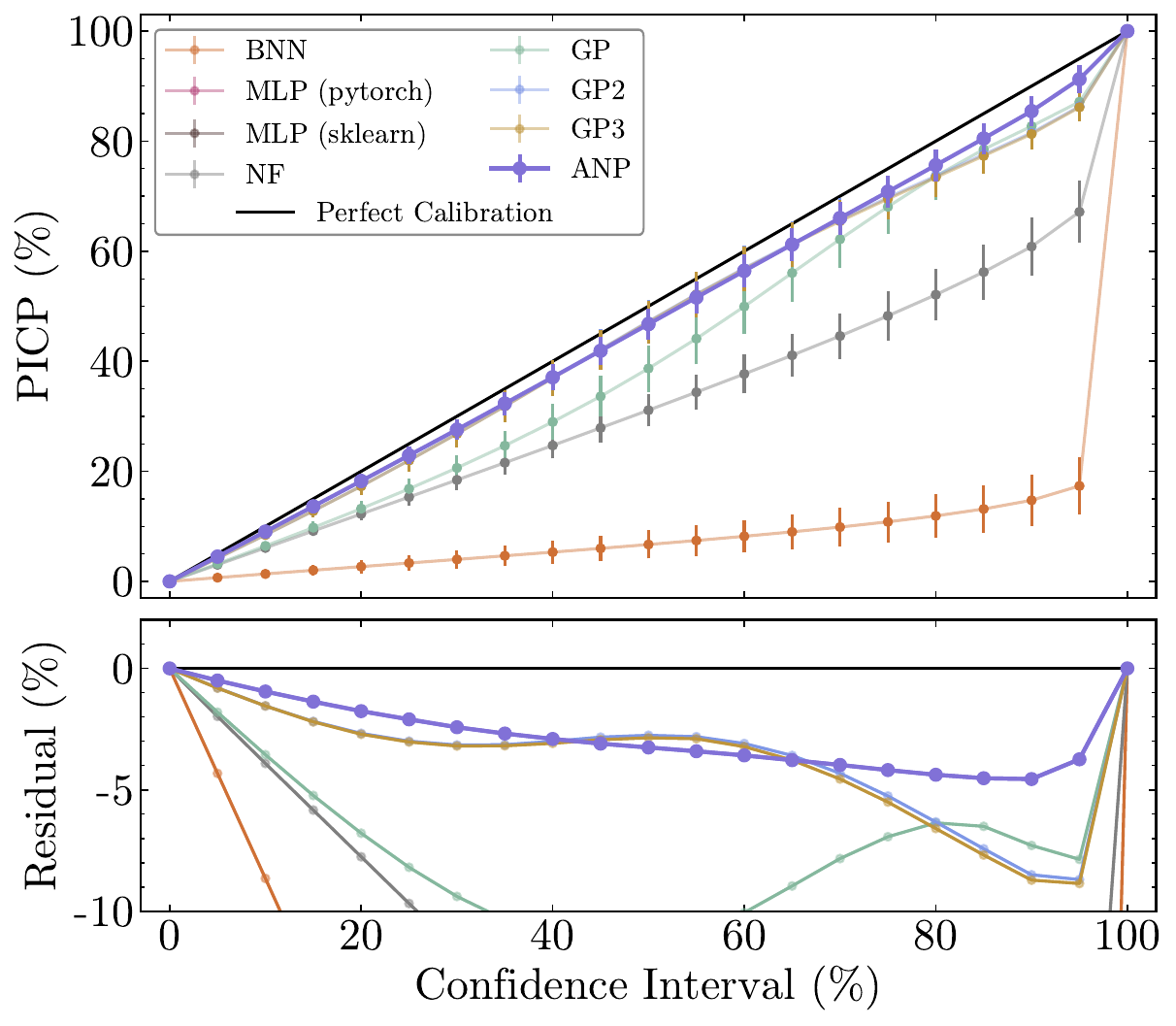}
    \caption{Results VI: Confidence calibration plot for ANP and all seven benchmark models considered in this work (see \autoref{sec:results}). The top panel displays the empirical coverage (PICP, see \autoref{subsec:results-probabilistic}) as a function of the predicted confidence interval, where the solid black diagonal line represents perfect statistical calibration. The bottom panel shows the calibration residuals (empirical PICP minus predicted confidence). All values are median-aggregated over 10 cadence realizations. 
    ANP achieves near-perfect calibration with residuals \textless 5\%. The BNN and MLP are drastically overconfident, while GP2 and GP3 are the closest to this ideal but still slightly worse than ANP. We note that the ANP's calibration residual decreases monotonically (almost linearly), making the ANP's uncertainties more trustworthy for further analysis.
    } 
    \label{fig:calibration}
\end{figure}

We note some differences in the ANP calibration based on the training strategy (\autoref{subsec:procedure}). 
\autoref{fig:calibration} shows the PICP as a function of confidence interval. Not only does the ANP have the best calibration, with residuals $<5\%$ from a perfect calibration, but the ANP calibration curve is highly stable, monotonically decreasing up to 95\% confidence. This near-linear behavior is preferable to an erratic, oscillating calibration curve as it indicates a predictable relationship between the model confidence and empirical coverage. In contrast, the GPs' performance, while overall showing comparable calibration residuals, fluctuates and worsens more abruptly at higher confidence levels. This indicates the ANP uncertainties are trustworthy and reliable.

To jointly evaluate calibration and sharpness in a single metric, \citealt{gneitingProbabilisticForecastsCalibration2007} recommends the use of \emph{proper scoring rules}, where a `strictly proper' scoring rule is one that is maximized (or minimized) if and only if the predicted probability distribution perfectly matches the ground truth distribution (see \citealt{gneitingStrictlyProperScoring2007, pion-vasquez-2025} for a review). Our next two metrics then are two such metrics: the Negative Log Predictive Density (NLPD) and the Continuous Ranked Probability Score (CRPS). The NLPD heavily penalizes models when the model probability of a ground truth flux value is low, but is known to be sensitive to outliers. The CRPS is a probabilistic generalization of the MAE, in the units of flux.

\begin{figure}
    \centering
    \includegraphics[height=0.75\linewidth]{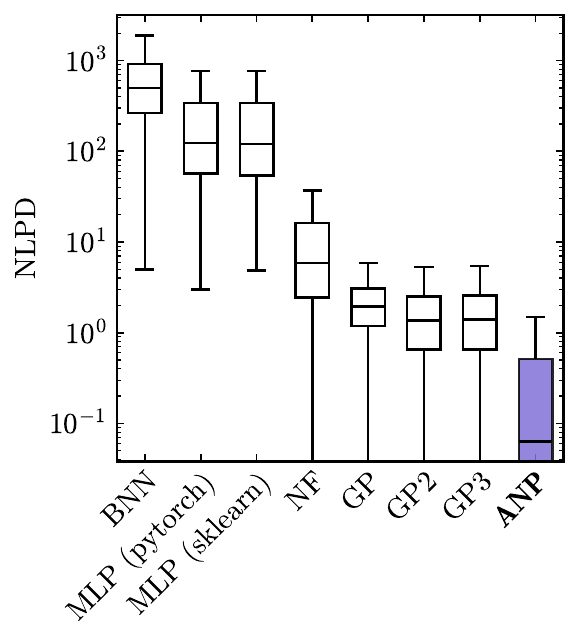}
    \includegraphics[height=0.75\linewidth]{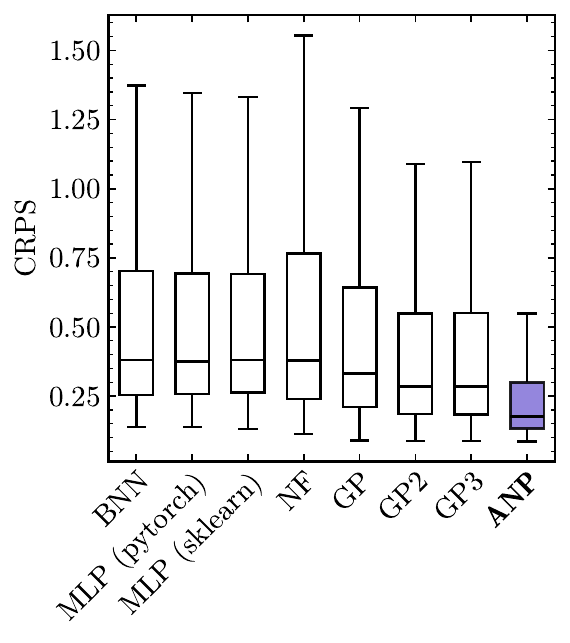}

    \caption{Results VII: Distribution of two probabilistic evaluation metrics, based on the \cite{gneitingProbabilisticForecastsCalibration2007} framework (lower is better). Like in \autoref{fig:result-regression}, all metrics are median-aggregated over 10 cadence realizations. (a) Negative Log Predictive Density (NLPD), a proper scoring rule that jointly penalizes both poor point predictions and miscalibrated uncertainties, plotted on the log-scale. (b) Continuous Ranked Probability Score (CRPS), a generalization of the Mean Absolute Error to probabilistic forecasts.}
    \label{fig:result-prob}
\end{figure}

\autoref{fig:result-prob} shows the distribution of the NLPD (upper) and CRPS (lower) scores for the different models. The ANP's NLPD values are over an order of magnitude better than the next-best GP models, and the ANP also has the lowest CRPS values on average, although the difference is less dramatic. Together, these imply that the ANP consistently makes predictions which are both sharp and centered accurately on the ground truth flux value (see also \autoref{app:prob} for an analysis of the Probability Integral Transform).

\astrover{

\section{Conclusions} \label{sec:conclusion}

Astrophysical light curves are characteristically sparse and irregular, posing a challenge for statistical and machine learning methods.
The upcoming Vera C. Rubin Observatory's LSST will generate millions of transient alerts nightly, presenting an unprecedented data-processing challenge. With a highly constrained survey strategy, designed to achieve multiple goals at once \citep{ivezicLSSTScienceDrivers2019, PSTN-056, Bianco_2022}, the LSST cadence will typically collect one-three observations within a night in different filters, and revisit the field a few days later, leaving gaps of considerable size compared to the typical evolutionary time scale of most transients (days, weeks, months). To maximize science with this data, time-domain astronomy requires models capable of interpolating, or reconstructing, sparse, irregularly sampled, multi-band light curves. Traditional interpolation methods, such as Gaussian Processes (GPs), scale poorly ($\mathcal{O}(N^3)$) and require per-object fitting, making them computationally prohibitive to use on datasets as large as the LSST alert stream. To address this, we proposed the use of Attentive Neural Processes (ANPs) as fast, data-driven, universal light curve interpolators.

We developed an implementation of the ANP specifically designed for inference on sparse, multi-band light curves typical of astronomical surveys. Our ANP, which incorporates attention, is entirely data-driven and morphologically agnostic, serving as a highly parallelizable solution that can work at the microsecond speeds necessary for the live alert stream. By leveraging meta-learning, the ANP allows for amortized inference. This moves the computational cost from inference to training, and enables inference-time interpolation at $\sim 10^{-6}$ seconds per light curve on a single GPU -- several orders of magnitude faster than GPs and other neural benchmark models (\autoref{sec:results}). 

We tested the performance of the ANP at light curve reconstruction against the performance of seven benchmark models commonly used in astrophysical research following a well-established paradigm \citep{demianenko_understanding_2023}. We found the ANP to be dramatically computationally advantageous, as well as superior to all the models tested under a total of 15 distinct performance metrics, some specifically designed for this work, others taken from both the astrophysical and statistical literature. Whether measuring raw regression quality (\autoref{subsec:results-regression}), the recovery of astrophysical features (\eg, peak-time and magnitude, \autoref{subsec:results-peak}), or rigorous probabilistic calibration (\eg, CRPS, PIT histograms, \autoref{subsec:results-probabilistic}), the ANP consistently outperforms all tested benchmark models. In addition, the ANP is highly calibrated, avoiding both the over-conservative uncertainties of GPs and the catastrophic overconfidence typical of standard neural networks, making its probabilistic outputs highly reliable.

\subsection{Limitations} \label{sec:limitations}

While our unoptimized model successfully demonstrates the viability of the meta-learning framework for time-domain astronomy, there are several limitations to the current implementation.

First, in this proof-of-concept study, we trained and tested our models on synthetic data derived from the PLAsTiCC \citep{hlozekResultsPhotometricLSST2023, plasticcmodelersLibrariesAmpRecommended2022} models. These simulations are widely accepted as highly effective at representing the physical morphologies and noise distributions expected from LSST. We train on simulations so as to provide dense-cadence multi-band data to the ANP model, allowing it to learn the underlying temporal and wavelength correlations -- data that is largely unavailable in current real-world surveys. However, translating to real light curves will require some engineering adaptations. Deploying a production-ready version for the live LSST alert stream will naturally require engineering the optimal architecture through component-wise ablations and hyperparameter tuning deferred in this study. The immediate next step is to deploy and evaluate this framework on true observational datasets, such as those from the Zwicky Transient Facility (ZTF) and early LSST alerts.

Second, the four astronomical adaptations (band-as-coordinate, dynamic slicing, temporal re-zeroing, stratified sampling) were chosen for documented engineering reasons (ragged-tensor handling, GPU memory, translation-equivariance, class imbalance) rather than empirical sweeps. We also did not do a thorough architectural \& hyperparameter search, which we expect to improve performance.

Third, our implementation of ANP did not take the observational uncertainty as input. Observational uncertainties are complex but exceptionally well understood in most astronomical surveys and extremely valuable in inference. Relatively small modifications of our architectures would allow the integration of observational uncertainties in the context points and we leave this to future work. 

Finally, while our dataset contained different explosive transient classes, the ability of ANP to reconstruct light-curves driven by radically different physical processes remains to be demonstrated: can the same ANP correctly model supernovae, variable stars and transiting planets? 

\subsection{Future Directions} \label{sec:future}

We establish that the Neural Process family provides a robust, natural foundation for astronomical time-series analysis. Looking forward, the value of this family of models extends beyond simple interpolation. Accurate forecasting could be useful for prioritizing follow-up resources, while the reconstructions themselves can be directly served to existing machine learning classification heads, mitigating the challenges of missing data and potentially improving classifier performance. 

Furthermore, because of the ANP's natural ability to learn a robust, continuous representation from irregular multi-band data, its latent space ($\vec{\bm{z}}$, \autoref{sec:methods}) encodes a rich, dimensionally reduced summary of the light curve. We are yet to exploit the ANP latent space for inference: this latent representation may serve as an exceptionally powerful feature space for downstream scientific tasks, including physical parameter estimation, similarity searches, and the identification of true novelties (`unknown unknowns') in the imminent LSST data stream to accelerate discoveries in time-domain astrophysics.

Finally, while we have justified our choice of ANP (\autoref{sec:methods} and \autoref{sec:conclusion}), other NP flavors could be implemented for multiband light curves and should be compared with ANP as well as neural ODE methods such as SELDON \citep{wuSELDONSupernovaExplosions2026}, once implemented in a class-agnostic framework.

\begin{acknowledgments}
SC acknowledges support received from the University of Delaware Doctoral Fellowship of Excellence and the NASA FINESST program, Grant 80NSSC25K0312. FBB is supported in part by NSF AST Award Numbers 2511639 and 2308016. AAM is supported in part by NASA Grant 80NSSC24M0020. 

SC and FBB are grateful to Andjelka~B.~Kova{\v{c}}evi{\'c} and Dragana~Ili{\'c} for insightful conversations on neural processes; Emille Ishida and David Jones, along with the PLAsTiCC Team, for the simulated data. SC is also grateful to Konstantin Malanchev, Oleksandra Razim and members of FASTLab for helpful discussions which helped shape this work.

The authors acknowledge the support of the Vera C. Rubin Legacy Survey of Space and Time Science Collaborations\footnote{\url{https://www.lsstcorporation.org/science-collaborations}} and particularly of the Transient and Variable Star Science Collaboration\footnote{\url{https://lsst-tvssc.github.io/}} (TVS SC) that provided opportunities for collaboration and exchange of ideas and knowledge.

\end{acknowledgments}

\vspace{5mm}

\software{\texttt{SNANA} \citep{snana}, \texttt{fulu} \citep{demianenko_understanding_2023}, \texttt{rubin\_sim} \citep{peteryoachimLsstRubin_simV2612026}, \texttt{LightCurveLynx} \citep{lightcurvelynx}, \texttt{Keras} \citep{chollet2015keras}, \texttt{TensorFlow} \citep{tensorflow2015-whitepaper}, \texttt{jax} \citep{jax}, \texttt{PyTorch} \citep{pytorch}, \texttt{NumPy} \citep{numpy}, \texttt{Jupyter} \citep{jupyter}, \texttt{matplotlib} \citep{matplotlib}, \texttt{seaborn} \citep{seaborn}, \texttt{scikit-learn} \citep{scikit-learn}, \texttt{SciPy} \citep{scipy}, \texttt{pandas} \citep{pandas,pandas2}, \texttt{polars} \citep{polars}, \texttt{tqdm} \citep{tqdm}, and \texttt{Python3} \citep{python3}.}

\bibliography{refs}{}
\bibliographystyle{aasjournal}

\clearpage

\appendix
\section{Summary statistics} \label{app:summary}

\autoref{tab:metrics_summary} summarizes the performance of ANP and all comparison models across all metrics described in \autoref{subsec:reconstruction-performance} (see also \autoref{fig:winrate comparison}). ANP is the top-performing model (or equal to the top-performing model within uncertainties) across all metrics. Note that the better performance of BNN on MPIW is in fact problematic: it indicates the model has high precision, but also low accuracy (poor PICP, see \autoref{subsec:results-probabilistic}). Similarly, note that GPs' performance on forecasting peak-time is driven by an inherent bias (see \autoref{subsec:results-peak}).

\section{Probability Integral Transform}\label{app:prob}
While the summary statistics described in \autoref{subsec:reconstruction-performance} are useful for ranking models, they do not provide details on how a model's calibration fails. For this, we look at the Probability Integral Transform (PIT). The PIT is defined as the value of the model's predicted cumulative distribution function (CDF) evaluated at the ground truth flux value. For a perfectly calibrated model, the true flux values should fall randomly and uniformly anywhere within the predicted distribution, resulting in PIT values that follow a standard Uniform(0, 1) distribution. We plot these as histograms in \autoref{fig:pit}. To quantify the deviation from uniformity, we also calculate the Kolmogorov-Smirnov (KS) statistic, which measures the maximum absolute distance between the empirical PIT and the ideal Uniform(0, 1) CDF.\footnote{We note none of these distributions, which include $>10^6$ data points each, are statistically consistent with a uniform distribution at the $>5\sigma$ level; we include the KS statistic as a metric of similarity, not a strict test of statistical consistency.}

We see that the MLPs and BNN exhibit severe U-shapes with very large spikes at the boundaries, indicative of their catastrophic overconfidence. The NF also follows a U-shape, but is an improvement toward the Uniform(0,1) ideal CDF, as evident from its KS value (0.174). The GPs are better calibrated overall (KS=0.134 for GP3), but exhibit a left-skew in their PIT, along with spikes near the boundaries. This implies that the true flux values frequently fall in the lower percentiles (0.1 - 0.4) of the predicted distribution, indicating the GPs display a positive bias, systematically over-predicting the mean flux, but fail to bound outliers (spikes at 0 and 1).

Meanwhile the ANP produces the flattest histogram, close to the Uniform(0,1) ideal, with a vastly lower KS statistic (0.048). While it still exhibits spikes near the boundaries, and a shallow dip near 0.8, these deviations are much smaller than for the other benchmark models. We conclude that the ANP avoids the severe overconfidence of standard neural methods, while also avoiding the systematic bias displayed by GPs, and successfully learns to produce highly calibrated uncertainties.

\begin{figure*}
    \centering
    \includegraphics[width=0.95\linewidth]{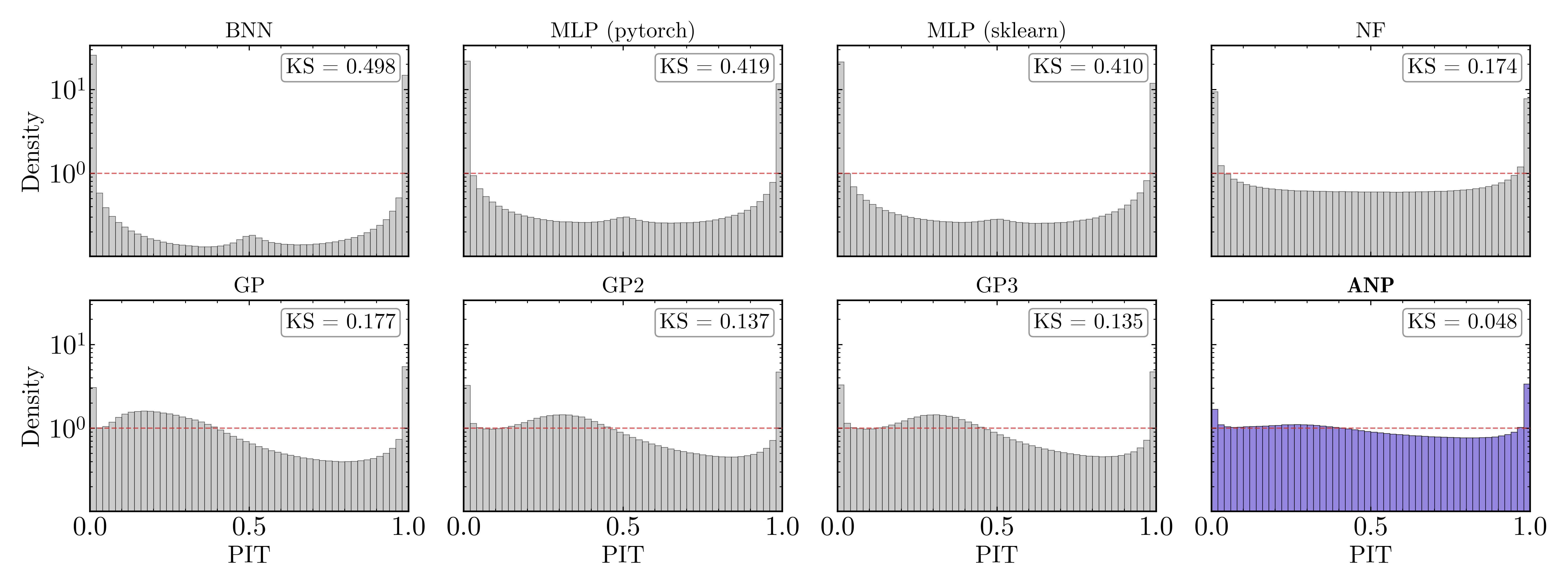} %
    \caption{Probability Integral Transform histograms (PIT, \citealt{gneitingProbabilisticForecastsCalibration2007}) for all eight models considered in this work. For a perfectly calibrated model, PIT values follow a Uniform(0, 1) distribution, producing a flat histogram (red dashed line at density = 1). PIT values are calculated for all points across all light curves and across all realizations. %
    The KS statistic (lower is better; derived from the Kolmogorov-Smirnov test) is indicated in each panel. The value denotes the maximum distance between the cumulative empirical and Uniform distributions.
    ANP vastly outperforms the other methods.}
    \label{fig:pit}
\end{figure*}

\section{ANP Performance for different astrophysical classes} \label{subsec:allresults-by-class}
A core motivation for using ANPs for interpolation is their ability to behave as a `universal' interpolator -- a single model that does not require the user to know the transient class a priori in order to select a kernel or functional form to model that class. To verify that an ANP trained on a variety of classes generalizes well across different physical phenomena, we break down the MSSE by broad physical class (see \autoref{sec:data}) in \autoref{fig:allresults-by-class}. %
The ANP performs well on reconstruction, including in some data subsets that combined classes with intrinsically diverse photometric properties (\eg, Core-Collapse Supernovae). %

We note that the overall reconstruction errors for Thermonuclear supernovae (primarily Type Ia) are higher and have higher variance across all models compared to the CC and SE classes. Given that thermonuclear supernovae are largely homogeneous, with light curves dominated by $^{56}$Ni evolution, this is initially surprising. We attribute this to an artifact of the underlying simulation data we used; the simulated SN Ia light curves baseline is typically $\sim400$ days, whereas the CC and SE simulations span a shorter 150--200 day window. SN Ia simulations are padded by long, zero-flux pre- and post-peak baselines. Because we construct our context sets using random LSST-like sampling, a significant fraction of the SN Ia realizations consist entirely of context points drawn from these information-poor regions. 
However, visual inspection of the light curves suggests that when context points are available near the peak, all models perform well, and the ANP consistently yields the lowest errors. Furthermore, in intermediate cases where only a few points lie on the rise/decline, the ANP significantly outperforms the benchmarks.

Finally, for the subclass containing CARTs and ILOTs, all models perform comparably.  
These transients consist of different physical processes that lead to light curve morphologies and SEDs that can be different from one another, and more distinct from that of the other sub-classes of transients. Thus, in comparison, the ANP has naturally seen fewer examples of these subclasses during training.}

\begin{figure*}
    \centering
    \includegraphics[width=\linewidth]{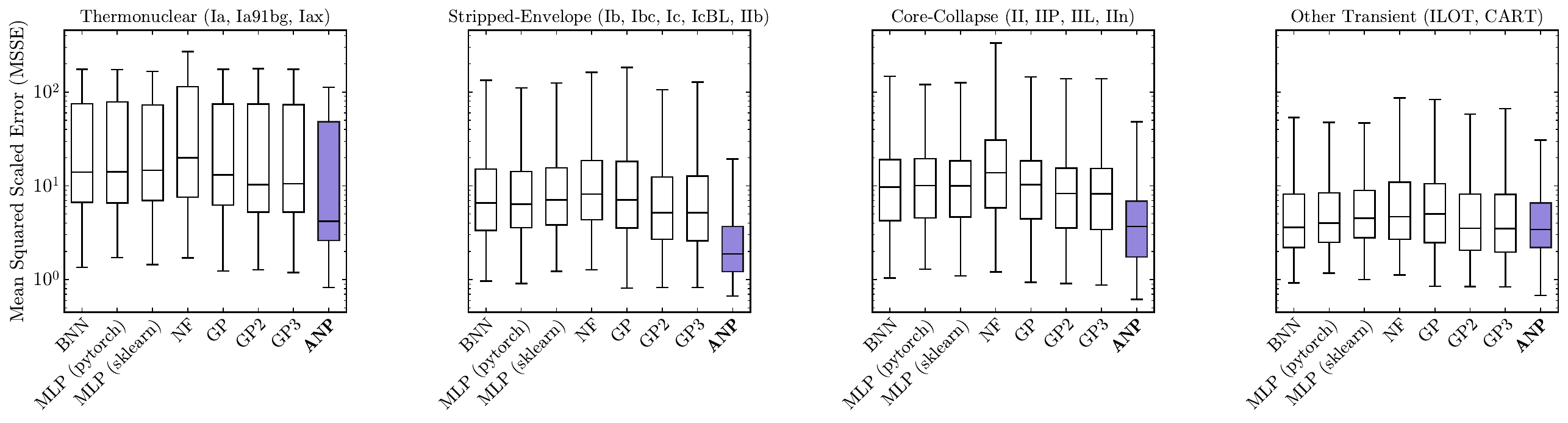}
    \caption{Results VIII: Distribution of the Mean Squared Scaled Error as a function of the physical class; SN-like transients are modelled most effectively by ANP. For thermonuclear SNe, which are a more homogeneous set of SNe, ANP still outperforms other models but with a less dramatic difference. For Other transients, which include ILOT and CART (see \autoref{sec:data}), the ANP predictions are statistically as good as those of other models.}
    \label{fig:allresults-by-class}
\end{figure*}

\begin{longrotatetable}
\begin{deluxetable*}{lcccccccc}
\tablecaption{Summary of model performance across regression, uncertainty-aware, and peak-estimation metrics. \label{tab:metrics_summary}}
\tablehead{
    \colhead{Metric} & \colhead{BNN} & \colhead{MLP (pytorch)} & \colhead{MLP (sklearn)} & \colhead{NF} & 
    \colhead{GP} & \colhead{GP2} & \colhead{GP3} & \colhead{\textbf{ANP}}
}
\startdata
\cutinhead{Standard Regression}
MSE & $0.46^{+1.36}_{-0.29}$ & $0.46^{+1.34}_{-0.28}$ & $0.48^{+1.30}_{-0.29}$ & $0.60^{+1.98}_{-0.39}$ & $0.49^{+1.34}_{-0.30}$ & $0.37^{+1.04}_{-0.23}$ & $0.37^{+1.07}_{-0.24}$ & $\mathbf{0.18}^{+0.50}_{-0.10}$ \\
MSSE & $7.95^{+11.28}_{-4.28}$ & $8.06^{+11.30}_{-4.08}$ & $8.48^{+10.71}_{-4.24}$ & $10.41^{+17.76}_{-5.64}$ & $8.55^{+11.24}_{-4.52}$ & $6.60^{+9.16}_{-3.44}$ & $6.56^{+9.36}_{-3.46}$ & $\mathbf{3.01}^{+3.53}_{-1.38}$ \\
MAPE & $748^{+774}_{-272}$ & $784^{+758}_{-279}$ & $729^{+660}_{-276}$ & $1126^{+1316}_{-435}$ & $1089^{+1208}_{-470}$ & $888^{+968}_{-374}$ & $878^{+963}_{-371}$ & $\mathbf{294}^{+220}_{-82}$ \\
MASE & $2.21^{+1.32}_{-0.64}$ & $2.25^{+1.26}_{-0.60}$ & $2.27^{+1.23}_{-0.59}$ & $2.55^{+1.63}_{-0.75}$ & $2.42^{+1.61}_{-0.79}$ & $2.09^{+1.29}_{-0.63}$ & $2.08^{+1.29}_{-0.63}$ & $\mathbf{1.37}^{+0.61}_{-0.27}$ \\
\cutinhead{Uncertainty-Aware Regression}
$\chi^2 \ (\times 10^3)$ & $787^{+801}_{-368}$ & $210^{+357}_{-114}$ & $205^{+366}_{-111}$ & $10.9^{+16.9}_{-5.5}$ & $2.98^{+1.57}_{-0.87}$ & $2.28^{+1.59}_{-0.70}$ & $2.32^{+1.60}_{-0.71}$ & $\mathbf{1.02}^{+0.34}_{-0.16}$ \\
NRMSE$_\mathrm{p}$ & $31.3^{+11.3}_{-8.4}$ & $15.6^{+10.3}_{-4.9}$ & $15.4^{+10.2}_{-4.9}$ & $3.61^{+2.14}_{-1.06}$ & $1.90^{+0.42}_{-0.27}$ & $1.68^{+0.44}_{-0.27}$ & $1.69^{+0.44}_{-0.27}$ & $\mathbf{1.13}^{+0.14}_{-0.07}$ \\
NRMSE$_\mathrm{o}$ & $4.58^{+4.23}_{-1.79}$ & $4.61^{+4.15}_{-1.70}$ & $4.70^{+4.01}_{-1.68}$ & $5.27^{+5.34}_{-2.07}$ & $4.68^{+4.09}_{-1.75}$ & $4.10^{+3.68}_{-1.53}$ & $4.13^{+3.73}_{-1.59}$ & $\mathbf{2.82}^{+2.46}_{-0.94}$ \\
NLPD & $496^{+425}_{-230}$ & $123^{+219}_{-66}$ & $120^{+221}_{-66}$ & $5.84^{+10.42}_{-3.40}$ & $1.94^{+1.12}_{-0.76}$ & $1.36^{+1.15}_{-0.71}$ & $1.40^{+1.17}_{-0.75}$ & $\mathbf{0.06}^{+0.45}_{-0.21}$ \\
CRPS & $0.38^{+0.43}_{-0.15}$ & $0.38^{+0.44}_{-0.16}$ & $0.39^{+0.45}_{-0.17}$ & $0.40^{+0.63}_{-0.18}$ & $0.34^{+0.33}_{-0.13}$ & $0.29^{+0.31}_{-0.11}$ & $0.29^{+0.31}_{-0.11}$ & $\mathbf{0.18}^{+0.17}_{-0.05}$ \\
$\text{PICP}_{0.95}$ (\%) & $14.2^{+9.3}_{-4.8}$ & $31.0^{+4.5}_{-4.5}$ & $31.7^{+4.2}_{-4.0}$ & $69.6^{+5.2}_{-7.4}$ & $88.3^{+2.2}_{-3.3}$ & $87.1^{+2.8}_{-3.6}$ & $86.9^{+2.8}_{-3.6}$ & $\mathbf{92.7}^{+1.6}_{-2.6}$ \\
$\text{MPIW}_{0.95}$ & $\mathbf{0.10}^{+0.11}_{-0.04}$ & $0.22^{+0.06}_{-0.04}$ & $0.22^{+0.05}_{-0.04}$ & $0.78^{+0.19}_{-0.07}$ & $1.81^{+1.43}_{-0.66}$ & $1.44^{+1.27}_{-0.46}$ & $1.43^{+1.30}_{-0.49}$ & $1.01^{+0.47}_{-0.19}$ \\
\cutinhead{Peak-Estimation (days / mag)}
$|\Delta t_\mathrm{peak}|$  & $27^{+20}_{-12}$ & $28^{+20}_{-13}$ & $27^{+20}_{-12}$ & $47^{+26}_{-25}$ & $13^{+8}_{-5}$ & $12^{+8}_{-4}$ & $13^{+7}_{-5}$ & $\mathbf{8}^{+6}_{-3}$ \\
$|\Delta m_\mathrm{peak}|$  & $0.25^{+0.23}_{-0.11}$ & $\mathbf{0.24}^{+0.20}_{-0.11}$ & $0.24^{+0.19}_{-0.10}$ & $0.38^{+0.30}_{-0.19}$ & $0.35^{+0.26}_{-0.17}$ & $0.34^{+0.27}_{-0.16}$ & $0.34^{+0.27}_{-0.16}$ & $0.25^{+0.22}_{-0.10}$ \\
$|\Delta t_\mathrm{peak}|$ (per band) & $58.3^{+13.5}_{-14.0}$ & $59.0^{+14.0}_{-14.0}$ & $57.7^{+15.0}_{-15.7}$ & $62.7^{+17.0}_{-15.7}$ & $45.2^{+14.7}_{-13.2}$ & $48.0^{+14.0}_{-15.0}$ & $47.4^{+13.4}_{-14.8}$ & $\mathbf{18.2}^{+15.5}_{-9.0}$ \\
$|\Delta m_\mathrm{peak}|$ (per band) & $0.31^{+0.20}_{-0.12}$ & $0.30^{+0.18}_{-0.11}$ & $0.32^{+0.17}_{-0.11}$ & $0.35^{+0.23}_{-0.14}$ & $0.34^{+0.19}_{-0.12}$ & $0.31^{+0.18}_{-0.10}$ & $0.31^{+0.18}_{-0.11}$ & $\mathbf{0.23}^{+0.15}_{-0.07}$ \\
Forecasting $|\Delta t_\mathrm{peak}|$  & $27^{+25}_{-13}$ & $28^{+21}_{-12}$ & $28^{+20}_{-12}$ & $61^{+33}_{-25}$ & $14^{+14}_{-6}$ & $14^{+12}_{-6}$ & $14^{+12}_{-6}$ & $\mathbf{7}^{+4}_{-2}$ \\
Forecasting $|\Delta t_\mathrm{peak}|$ (per band) & $73.7^{+16.0}_{-16.0}$ & $74.2^{+17.3}_{-16.2}$ & $71.7^{+17.7}_{-16.7}$ & $89.5^{+29.5}_{-19.2}$ & $48.7^{+16.7}_{-15.8}$ & $43.7^{+19.7}_{-16.7}$ & $44.3^{+20.2}_{-16.7}$ & $\mathbf{16.8}^{+15.5}_{-7}$ \\
\enddata
\tablecomments{All metrics are reported as the median (Q2), with sub- and superscripts denoting the distance to the 25th (Q1) and 75th (Q3) percentiles, respectively. Bold values indicate the best-performing model for each metric.}
\end{deluxetable*}
\end{longrotatetable}

\end{document}